\documentclass[a4paper,11pt]{article}
\pdfoutput=1 
\usepackage{jheppub}  
\usepackage{graphicx,color}
\usepackage{amsmath}
\usepackage{autobreak}
\usepackage[normalem]{ulem}
\usepackage{multirow}

\allowdisplaybreaks
\newcommand{\ben}{\begin{enumerate}}
\newcommand{\een}{\end{enumerate}}
\newcommand{\beq}{\begin{equation}}
\newcommand{\eeq}{\end{equation}}
\newcommand{\bal}{\begin{align}}
\newcommand{\eal}{\end{align}}

\newcommand{\bea}{\begin{eqnarray}}
\newcommand{\eea}{\end{eqnarray}}

\def\z#1{{\zeta_{#1}}}

\newcommand{\Ca}{C_A}
\newcommand{\Cf}{C_F}
\newcommand{\nf}{n_F}

\newcommand{\df}{{\rm d}}
\def\Dm1{{{\delta(1-z)}}}

\def\nf{{n^{}_{\! f}}}

\def\g0#1DY{{g_{0#1}^{DY}}}

\newcommand{\Lqr}{L_{qr}}
\newcommand{\Lfr}{L_{fr}}

\def\LogmW1{{{\ln (1-\omega)}}}

\newcommand{\as}{a_s}
\newcommand{\muf}{\mu_F}
\newcommand{\mur}{\mu_R}
\newcommand{\mz}{m_z}

\newcommand{\eq}[1]{Eq.\ (\ref{#1})}
\newcommand{\fig}[1]{Fig.\ [\ref{#1}]}
\newcommand{\tab}[1]{Tab.\ [\ref{#1}]}

\title{{\boldmath 
Threshold resummation for $W$-boson pair production at NNLO+NNLL}}

\author[a]{Pulak Banerjee,}
\author[b]{Chinmoy Dey,}
\author[b]{M. C. Kumar }
\author[b]{and Vaibhav Pandey}
\affiliation[a]{Istituto Nazionale di Fisica Nucleare, Gruppo collegato di Cosenza,
I-87036 Arcavacata di Rende, Cosenza, Italy }
\affiliation[b]{Department of Physics, 
	Indian Institute of Technology Guwahati, Guwahati-781039, India}
\emailAdd{pulak.banerjee@lnf.infn.it}
\emailAdd{d.chinmoy@iitg.ac.in}
\emailAdd{mckumar@iitg.ac.in}
\emailAdd{vphiitg@iitg.ac.in }
\abstract{
We present results for threshold resummation of the invariant mass distribution, for on-shell production of a pair of $W$ bosons at next-to-next-to-leading order +  next-to-next-to-leading logarithmic(NNLO+NNLL) accuracy in QCD. 
Owing to its sensitivity to the self-interactions between gauge bosons, this process is important to investigate at the energies of the Large Hadron Collider(LHC). 
We achieve this resummation by exploiting the factorization properties of the soft and virtual parts of the partonic cross-section.
Our analysis has been carried out for the invariant mass distribution up to $Q$ = 2500 GeV.
At this highest $Q$ we find that, for 13.6 TeV LHC, the NNLL resummation enhances the NNLO cross-sections by about $6.3\%$ and reduces the conventional scale uncertainties from $6.8\%$ at NNLO to $4.1\%$ at NNLO+NNLL. We also estimate the intrinsic uncertainties due to the non-perturbative parton distribution functions at the highest perturbative order, for both fixed-order and resummed results, to be around $3\%$ for $Q \sim$ 2000 GeV. 
}

\begin{document}

\keywords{Resummation, perturbative QCD, LHC}
\maketitle

\section{Introduction} \label{sec:introduction}
The Standard Model(SM) of particle physics has been successful in explaining intricacies inherent in the world around us, like the electron's magnetic moment, discovery of the Higgs boson \cite{ATLAS:2012yve,CMS:2012qbp}.
Yet it has many shortcomings, for example the origin of dark matter, matter-antimatter asymmetry in our universe, to name a few. 
Stringent limits have been put on physics beyond the Standard Model.
Production of an intermediate charged particle $X$ decaying to a W boson and a photon has been studied and no significant excess above the predicted background has been found \cite{CMS:2024ndg}.
Similarly, no new physics in high-mass diphoton events has been observed \cite{CMS:2024nht}. 
At the same time, collider experiments have measured fundamental parameters of the SM with extreme precision. One of the finest example is the precise measurement of the $W$ boson mass performed by LHC~\cite{CMS:2024nht} and by the CDF collaboration at the Tevatron~\cite{CDF:2013dpa}.
The importance of such precise measurements lies in the fact that any deviation from the SM value can signal the presence of beyond the SM(BSM) physics. 
For example, BSM physics could modify the $W$ boson mass via quantum loop effects \cite{Heinemeyer:2013dia}.
Precise SM theoretical predictions from the perturbative side have played a major role in $W$ boson mass measurements. \cite {CMS:2024lrd}.

It becomes important to precisely calculate observables in SM, up to high accuracy in perturbation theory. 
The production of a pair of massive gauge bosons at the LHC is an important process that has been studied extensively, both theoretically and experimentally. 
The process can also be used for testing the SM electroweak symmetry-breaking mechanism as well as in the study of fundamental weak interactions among elementary particles. 
This process is important, owing to the large cross-section for $W^{+} W^{-} $ (henceforth we call it $W$-boson pair) as compared to $ZZ$ or $W^{+} Z$ process, coupled with the inherent challenge in the final state mass reconstruction. 
Measurement of the $W$-boson pair production cross-sections has been done at CMS \cite{CMS:2024hey} as well as at ATLAS \cite{ATLAS:2022jat}. 
Significant progress has been made from the theoretical side, too. 
The leading order cross-section was calculated decades ago \cite{Brown:1978mq}. 
The next-to-leading order(NLO) corrections for this process were reported in \cite{Ohnemus:1991kk,Frixione:1993yp}. 
The helicity amplitudes at the NLO level\cite{Dixon:1998py} have enabled the calculation of the total production cross-section \cite{Campbell:1999ah}. 
The NLO results, including the full lepton decay correlations in the narrow-width approximation, can be found in \cite{Dixon:1999di}. 
The NLO corrections for $W$-boson pair in association with 1-jet was performed in  \cite{Dittmaier:2007th,Campbell:2007ev,Dittmaier:2009un,Hamilton:2016bfu}; for 2-jet results see \cite{Melia:2011dw,Greiner:2012im}; for 3-jets the results are available in \cite{FebresCordero:2015kfc}.
Inclusion of resummation effects at next-to-next-to leading log(NNLL) along with NLO corrections, for transverse momentum has been presented in \cite{Grazzini:2005vw,Meade:2014fca,Wang:2013qua}. 
For jet-veto resummation see \cite{Jaiswal:2014yba}. 
Electroweak effects have been studied in \cite{Bierweiler:2012kw,Baglio:2013toa,Billoni:2013aba}. 
For an effective field theory analysis at NLO, see \cite{Baglio:2017bfe,Baglio:2018bkm}. 
Polarisation fractions of massive gauge bosons serve as an important test of the symmetry-breaking mechanism in the SM. 
Polarised $W$-boson pair production at NLO and NNLO QCD is available in \cite{Denner:2020bcz} and \cite{Poncelet:2021jmj} respectively.

Precision studies entail going beyond NLO. The inclusive production of a on-shell $W$-boson pair in NNLO QCD was done in \cite{Gehrmann:2014fva}, where the residual perturbative uncertainty was at $3\%$. 
The NNLO helicity amplitudes for two off-shell bosons was calculated in \cite{Caola:2014iua,Gehrmann:2015ora}. 
The NNLO fiducial cross-sections and distributions for $W$-bosons pair are available \cite{Grazzini:2016ctr} decaying to four leptons. 
The leptonic decays lead to a decrease of the total cross-section as compared to the on-shell $W$-boson pair production. 
Upon including NLO EW corrections along with NNLO QCD, precise predictions are available \cite{Grazzini:2019jkl}. 
For fiducial cross-sections($\sigma$) at $\sqrt{s} = 13$ TeV, the NLO EW corrections for $W$-boson pair production is about -2.1\% compared to LO; $\sigma_{\text{NNLO QCD} + \text{NLO EW}}$ is about - 1.2\% compared to NNLO QCD; $\sigma_{\text{NNLO QCD} \bigotimes \text{NLO EW}}$ is about -2\% compared to NNLO QCD. 
For differential distributions, at $Q=1000$ GeV, the cross-sections for $\text{NNLO QCD} \bigotimes \text{NLO EW}$ and $\text{NNLO QCD} + \text{NLO EW}$ are about $-5\%$ compared to NNLO QCD \cite{Grazzini:2019jkl}. 
Planar master integrals at NNLO for $W$-boson pair production via top quark loop, in quark-antiquark annihilation channel, are now available \cite{He:2024iqg}. 
The loop induced gluon-fusion contributions which starts at $\mathcal{O}(a_s^2)$ at LO are also known \cite{Campbell:2011bn,Glover:1988fe,Binoth:2005ua,Binoth:2006mf}.
Two-loop helicity amplitudes for production of two off-shell vector bosons via gluon-fusion are available \cite{Caola:2015ila,vonManteuffel:2015msa}, which were later used to calculate the full NLO QCD corrections to $g g \rightarrow{W^{+}W^{-}}$ process~\cite{Caola:2015rqy}.

Significant progress has also been made in resumming large logarithmic contributions that arise in corner regions of phase space.
The transverse momentum resummation at NNLO+NNLL for a vector-boson pair was achieved in \cite{Grazzini:2015wpa}, where the dominant soft-gluon contributions at small $p_T$  have been resummed to all orders in perturbation theory.
Resummation for on-shell $W$-boson pair production with a jet-veto up to partial N$^3$LL + NNLO accuracy has been studied \cite{Dawson:2016ysj}. 
The logarithms resummed were the ratio of the transverse momentum of the vetoed-jet and the invariant mass of the final state. 
Threshold resummation at NNLL supplemented with approximate NNLO results has been investigated in \cite{Dawson:2013lya}. 
Using a SCET formalism, the authors have shown that the scale uncertainties (for invariant mass up to $500$ GeV) are smaller than the corresponding fixed order ones, upon inclusion of threshold logarithms. 
Parton shower analysis at NNLO for $W$-boson pair production has been studied in \cite{Re:2018vac, Lombardi:2021rvg}; using a jet veto, resummation within the SCET approach has been done in \cite{Gavardi:2023aco}.

Currently fixed-order invariant mass distributions have been reported till around $1000$ GeV  \cite{Gavardi:2023aco}. 
These cross-sections suffer from scale uncertainties arising from unphysical renormalization and factorization scales. 
If we set these scales to be the same and vary them between $(Q/2, 2Q)$ where Q is the invariant mass of the final state on-shell $W$-boson pair, we observe that for Q $\sim 500$ GeV, the uncertainty is about $2\%$. 
However, for $Q=2500$ GeV, this scale uncertainty rises to about $6.8\%$. 
A major programme of calculating higher-order corrections lies in reducing these scale uncertainties. 
These high $Q$ regions can be important in searches for signals beyond the SM \cite{CMS:2024ndg,CMS:2024nht}.
The CMS  has invariant mass distribution measured up to $1000$ GeV region for $ZZ$ production \cite{CMS:2020gtj}. 
With the upcoming HL-LHC, the integrated luminosity is expected to reach $3000-4000$ fb$^{-1}$, which may allow more events in this TeV region with more statistical data. 
For the planned FCC-hh \cite{FCC:2018vvp}, the increase in parton fluxes along with the integrated luminosity of $20-30$ ab$^{-1}$ can enhance the invariant mass distributions by a few orders of magnitude. 
A similar kind of situation related to the measurement in high $Q$ regions was also seen in the Drell-Yan process. 
One of the earliest measurements by CDF, for the Drell-Yan differential cross-section, was done till 1000 GeV \cite{CDF:2001gjd}. 
A decade later, CMS reported invariant mass distribution up to $600$ GeV while comparing to the theoretical predictions \cite{CMS:2011hqo}, followed by ATLAS measurement till $1500$ GeV \cite{ATLAS:2013xny}. 
Currently, the Drell-Yan invariant mass is measured up to $3000$ GeV \cite{CMS:2018mdl}. 
The residual theoretical scale uncertainty is about 0.1\% at N$^3$LO + N$^3$LL accuracy, for $Q=3000$ GeV \cite{Das:2022zie}.
One of the main motivations of the present work is to minimise the theoretical uncertainties that are as large as $6.8\%$ in the high invariant mass region.
After performing the resummation, the seven-point scale uncertainties in the NNLO fixed-order results are found to be reduced for the high invariant mass region to as low as 4.1\%.
Together with the fixed-order distributions, our resummed results give an estimate of the missing higher-order contributions.
It is to be noted that the resummation for on-shell $ZZ$ production process up to NNLO+NNLL has recently been done in ~\cite{Banerjee:2024xdh}. 

Our paper is organised as follows: in sec \ref{sec:theory} we discuss the theoretical framework for the NNLL resummation to the $q\bar{q}$ initiated $W$-boson pair production process. In sec \ref{sec:numerics}, we discuss our results, and we conclude in sec \ref{sec:conclusion}.

%

\section{Theoretical Framework} \label{sec:theory}
The hadronic cross-section for $W$-boson pair production  
can be written in terms of its partonic counterpart as: 
\begin{align}\label{eq:had-xsect}
 \frac{d\,\sigma}{d\, Q}  =
\sum_{a,b= \{q, \bar{q}, g\}}\int_0^1 dx_1\int_0^1 dx_2 \,\,f_{a}(x_1,\mu_F^2)\,
f_{b}(x_2,\mu_F^2)
\int_0^1 dz~ \delta \left(\tau-z x_1 x_2 \right)
 \frac{d\,\hat\sigma_{ab}}{d\, Q} \,,
\end{align}
where $Q$ is the invariant mass of the pair of $W$-bosons.The hadronic and partonic threshold variables $\tau$
and $z$ are defined as
\begin{align}
\tau=\frac{Q^2}{S}, \qquad z= \frac{Q^2}{s} \,,
\end{align}
where $S$ and $s$ are the hadronic and partonic center of mass energies, respectively. Thus, $\tau$ and $z$ are related by $\tau = x_1 x_2 z$.
The calculational framework that we follow is the same as in \cite{Banerjee:2024xdh}.
The leading order parton level process has the generic form
\begin{align}
	q(p_1) + \bar{q}(p_2) \to W^{+}(p_3) + W^{-}(p_4).
	\label{eq:parton}
\end{align}
We define kinematical variables $ s,t $ and $ u $ through,
\begin{align}
	(p_1+p_2)^2 = & s\,,~~~
	(p_1-p_3)^2 =  t\,,~~~ 
	(p_2-p_3)^2 =  u~~~
{\rm and} ~~~
	s + t + u = 2 m_{w}^2,
\end{align}
where, $m_{w}$ is the mass of the W-boson.
We also define $x,y$ and $z$ dimensionless scaling variables as,
\begin{align}
    x &= \frac{1-b}{1+b}, \quad 
    y = -\frac{t}{m_{w}^2}, \quad
    z = -\frac{u}{m_{w}^2}
\end{align}
where, $ b = \sqrt{1-{4 m_{w}^2}/{s}}$.
The leading-order (LO) cross-section $\hat{\sigma}^{(0)}_{q \bar{q}}$
for $W$-boson pair production can be written as, 
\begin{align}
	\frac{d\hat{\sigma}^{(0)}_{q \bar{q}}}{dQ} = \frac{1}{2 s}\int d\text{PS}_2 ~\mathcal{M}_{(0,0)},
	\label{zzbornVP}
\end{align}
where $d\text{PS}_2$ is the 2-body integral measure and  $ \mathcal{M}_{(0,0)} $ is the born amplitude, which for the up quark initiated process is given below,
\begin{align}\label{eq:born}
		\mathcal{M}_{(0,0)} = & B_{f} \,  {\rm N}
                                \left(c_{tt}^{u}\mathcal{F}_{0}^{u} - c_{ts}^{u}\mathcal{J}_{0}^{u} + c_{ss}^{u}\mathcal{K}_{0}^{u} \right).
\end{align}
Here,
\begin{align}
    \mathcal{F}_{0}^{u}(x,y)  = \frac{4}{x y^2} &\bigg[ {(-4 x+4 y+4 x^2 y+4 y^2+3 x y^2+4 x^2 y^2+y^3+x^2 y^3-x y^4)} \nonumber \\
          & +  \epsilon \left\{8 x-8 y-8 x^2 y-4 y^2-4 x^2 y^2\right\} \nonumber \\
        & + \epsilon^2 \left\{ {4 y+4 x^2 y-4 x (1+y^2)} \right\} \bigg], \\
    \mathcal{J}_{0}^{u}(x,y)  = \frac{1}{x^2 y} & \bigg[ 4 \, m_{w}^2 (y (4+y)+x^4 y (4+y) - 2 x^2 (10 - 4 y+y^2) - x (8 - 11 y+y^3)  \nonumber \\ 
    & - x^3 (8 - 11 y+y^3)
    + \epsilon \big\{ -16 m_{w}^2 (y+x^4 y+x (-2+3 y)+x^3 (-2+3 y) \nonumber \\
    & - x^2 (5 - 2 y+y^2)) \big\} \bigg], \\
    \mathcal{K}_{0}^{u}(x,y)  = \frac{1}{x^3} & \bigg[ 2 m_{w}^4 (4+y+x^6 (4+y)+x^2 (-4+11 y) + x^4 (-4+11 y)-x (-7+y^2) \nonumber \\ 
       & -x^5 (-7+y^2) - 2 x^3 (13+5 y^2)) + \epsilon \big\{ -8 m_{w}^4 (1+2 x+2 x^5+x^6 \nonumber \\ 
       & + x^2 (-1+2 y) + x^4(-1+2 y)-2 x^3 (3+y^2)) \big\}
            \bigg].
\end{align}
where $\epsilon$ is the dimensional regularization parameter for dimension $d = 4 - 2 \epsilon$.
The $\mathcal{F}$, $\mathcal{J}$ and $\mathcal{K}$ expressions for the $d$-type quark can be easily written in terms of $u$-type as,
\begin{align}
 \mathcal{F}^{d}_{0}(x,z) =  \mathcal{F}^{u}_{0}(x,y), \quad
 \mathcal{J}^{d}_{0}(x,z) = -\mathcal{J}^{u}_{0}(x,y), \quad
 \mathcal{K}^{d}_{0}(x,z) =  \mathcal{K}^{u}_{0}(x,y).
    \label{eq:dquarkBORN}
\end{align}
The coefficients $c_{tt}$,$c_{ts}$ and $c_{ss}$ are,
\begin{align}
    c_{tt} &= \frac{e^4}{16 \sin{\theta_{\rm w}}^2}, \nonumber \\
    c_{ts} &= \frac{e^4}{4 s \sin{\theta_{\rm w}}} \left( \mathcal{Q} + 2 e_{z} g_{L} \frac{s}{s-m_{z}^{2}} \right) , \nonumber \\
    c_{ss} &= \frac{e^4}{s^2}\left[\{ \mathcal{Q} + e_z (g_{L}+g_{R}) \frac{s}{s-m_{z}^2} \}^{2} + \{e_{z}(g_{L}-g_{R})\frac{s}{s-m_{z}^2}\}^{2} \right]
\end{align}
where, for $u$-type ($\mathcal{Q} = 2/3$) quark,
\begin{align}
    g_{L}^{u} = \frac{1}{2 \sin{\theta_{\rm w}} \cos{\theta_{\rm w}}} \left\{ \frac{1}{2}-\frac{2}{3} \sin{\theta_{\rm w}}^{2}\right\}, \quad
    g_{R}^{u} = -\frac{1}{2 \sin{\theta_{\rm w}} \cos{\theta_{\rm w}}} \left\{ \frac{2}{3} \sin{\theta_{\rm w}}^{2} \right\}
\end{align}
and, for d-type ($\mathcal{Q} = -1/3$) quark,
\begin{align}
    g_{L}^{d}= \frac{1}{2 \sin{\theta_{\rm w}} \cos{\theta_{\rm w}}} \left\{ -\frac{1}{2}+\frac{1}{3} \sin{\theta_{\rm w}}^{2}\right\}, \quad
    g_{R}^{d} = -\frac{1}{2 \sin{\theta_{\rm w}} \cos{\theta_{\rm w}}} \left\{ \frac{1}{3} \sin{\theta_{\rm w}}^{2} \right\}.
\end{align}
Here, $e^2 = 4 \pi \alpha$ with $\alpha$ being the fine structure constant, $\theta_{\rm w}$ is the weak mixing angle, {$e_{z}\equiv \cos{\theta_{\rm w}}/\sin{\theta_{\rm w}}$ and  $\mz$ is the mass of $Z$-boson. 
In the prefactor of \eq{eq:born}, N  is the SU(N) color, and $B_{f}(=1/4 \times 1/9)$  originates from the average of spin and color configurations. To calculate the amplitudes, we have dealt with $\gamma_5$ in $d$-dimensions using Larin's prescription \cite{Larin:1993tq}.

At higher orders in QCD, the partonic cross-section receives quantum corrections from virtual loops and real emissions. In the threshold region, defined by the limit $z \to 1$, the available energy in the partonic center-of-mass frame is almost entirely transferred to the production of the final state system. This leaves minimal phase space for additional parton radiations, making soft-gluon effects particularly significant. The partonic cross-section in terms of $z$ can be written as:
\begin{align}
\frac{d\,\hat \sigma_{ab}}{d\, Q} =
	\frac{d\,\hat \sigma^{(0)}_{ab}}{d\, Q}\left(
	\Delta_{ab}^{\rm sv}\left(z,\mu_F^2\right)
	+ \Delta_{ab}^{\rm reg}\left(z,\mu_F^2\right)
\right) \,.
\end{align}
The terms $\Delta_{ab}^{\rm sv}$ are universal and depend only on the initial state partons. 
The SV term encapsulates all singular behaviour in the limit $z \to 1$ as well as the threshold logarithms.
The remaining regular terms, denoted by $\Delta_{ab}^{\rm reg}$, are process-dependent and include 
logarithmic terms beyond the threshold, as well as the regular terms
that are finite in the limit $z \to 1$.
Notably, only quark-antiquark and gluon-gluon initiated channels contribute to the SV term. 

The singular structure of the SV component is universal and arises from contributions such as virtual form factors~\cite{Moch:2005tm, Moch:2005id, Baikov:2009bg, Gehrmann:2010ue, Gehrmann:2014vha}, mass factorization kernels~\cite{Moch:2004pa, Vogt:2004mw, Blumlein:2021enk}, and soft-gluon emissions~\cite{ Sudakov:1954sw, Mueller:1979ih, Collins:1980ih, Sen:1981sd, Sterman:1986aj, Catani:1989ne, Catani:1990rp,Kidonakis:1997gm,Kidonakis:2003tx,Ravindran:2005vv, Ravindran:2006cg,Moch:2005ba,Laenen:2005uz,Kidonakis:2005kz,Idilbi:2006dg}.
The process-dependent virtual terms in the threshold limit are proportional to $\delta(1-z)$, whereas the universal terms are proportional to the plus distributions, ${\cal D}_i = [\ln^i(1-z)/(1-z)]_+$. These contributions can be resummed to all orders in perturbation theory using Mellin-space techniques,
where convolutions of terms become normal products.

In Mellin space, the resummed SV cross-section can be written as:
\begin{align}\label{eq:resum-partonic}
	\hat{\sigma}_N^{\text{N}^n\text{LL}} =
	\int_0^1 \df z ~ z^{N-1} \Delta^{\rm sv}_{ab}(z)
	\equiv g_0 \exp\left(\Psi^{sv}_{N}\right),
\end{align}
where $g_0$ is independent of the Mellin variable $N$, and the exponent $\Psi^{sv}_N$ resums the large logarithms $\ln^i N$. Up to next-to-next-to-leading logarithmic (NNLL) accuracy, $\Psi^{sv}_N$ takes the form:
\begin{align}\label{eq:gn}
	\Psi^{sv}_N = \ln(\bar{N})\,{g}_1(\bar{N}) + {g}_2(\bar{N}) + \as\,{g}_3(\bar{N}) + \cdots,
\end{align}
with $\bar{N} = N\,e^{\gamma_E}$.
The universal functions ${g}_i$ are well known and their explicit expressions can be found in~\cite{Moch:2005ba, Catani:2003zt,Banerjee:2018vvb}. The constant $g_0$ encodes the non-logarithmic contributions and can be expanded in terms of the strong coupling constant $a_{s}\left(=g_s^2/(16 \pi^2)\right)$:
\begin{align}\label{eq:g0}
	g_0 = 1 + \as \,g_{01} + \as^2\,g_{02} + \cdots,
\end{align}
The coefficients $g_{01}$ and $g_{02}$ are process-dependent and are derived using the method outlined in~\cite{Banerjee:2017cfc,Ahmed:2020nci}, and their expressions can be found in Appendix \ref{appendixa}. 
We have used our in-house \texttt{FORM} ~\cite{Ruijl:2017dtg} routines to compute the one-loop virtual corrections (see Appendix \ref{appendixa}), which are necessary for obtaining $g_{01}$. 
The computation of the $g_{02}$ coefficient, which is essential for NNLL accuracy, requires the one-loop ($\mathcal{M}_{(0,1)}$), one-loop squared ($\mathcal{M}_{(1,1)}$) and the two-loop ($\mathcal{M}_{(0,2)}$) amplitudes. The $\mathcal{M}_{(0,1)} $ as well as the $\mathcal{M}_{(1,1)}$ amplitudes are calculated using in-house codes. For the $\mathcal{M}_{(0,2)}$ amplitudes, we have used the public package VVamp \cite{Gehrmann:2015ora} 
to assemble the ultraviolet and infrared finite contributions, which were then used in $g_{0}$ expression \ref{eq:g0}.

The resummed cross-section in $z$-space can now be obtained by performing the inverse Mellin transformation of the resummed cross-section in Mellin space:
\begin{align}
	\frac{d\sigma^{\text{N}^n\text{LL}}}{d Q} =
	\frac{d\hat{\sigma}^{(0)}}{d Q}
	\sum_{a,b \in \{q,\bar{q}\}}
	\int_{c-i\infty}^{c+i\infty}
	\frac{dN}{2\pi i}
	\tau^{-N}
	f_{a,N}(\mu_F)\,
	f_{b,N}(\mu_F)\,
	\hat{\sigma}_N^{\text{N}^n\text{LL}} \,.
\end{align}

Here, following the minimal prescription~\cite{Catani:1996yz}, we choose the contour of integration in the complex plane to avoid the Landau pole at $N = \exp(1/(2\as \beta_0) - \gamma_E)$. We set $N = c + x\,e^{i\phi}$ with $c = 1.9$ and $\phi = 3\pi/4$, as in~\cite{Vogt:2004ns}.

The differential cross-section after matching to the fixed-order counterparts is given by:
\begin{align}\label{eq:matched}
	\frac{d\sigma^{\text{N}^n\text{LO}+\text{N}^n\text{LL}}}{d Q} &=
	\frac{d\sigma^{\text{N}^n\text{LO}}}{d Q} +
	\frac{d\hat{\sigma}^{(0)}}{d Q}
	\sum_{a,b \in \{q,\bar{q}\}}
	\int_{c-i\infty}^{c+i\infty}
	\frac{dN}{2\pi i}
	\tau^{-N}
	f_{a,N}(\mu_F)\,
	f_{b,N}(\mu_F) \nonumber \\
	&\times
	\left[
	\hat{\sigma}_N^{\text{N}^n\text{LL}} -
	\left.\hat{\sigma}_N^{\text{N}^n\text{LL}}\right|_{\text{tr}}
	\right].
\end{align}
Here, $\left.\hat{\sigma}_N^{\text{N}^n\text{LL}}\right|_{\text{tr}}$ denotes the truncation of the resummed cross-section at the $n^{\text{th}}$ order to avoid double counting. Mellin-space PDFs, $f_{a,N}$, are obtained via tools such as \\ \texttt{QCD-PEGASUS}~\cite{Vogt:2004ns}, or approximated using $z$-space PDFs as in~\cite{Catani:2003zt, Catani:1989ne}.

\section{Numerical Results}\label{sec:numerics}
In this section, we present the numerical results for the on-shell production of the 
$W$-boson pairs at the current LHC as well as the future high energy hadron colliders. Most of the results are presented for $\sqrt{S}=13.6$ TeV.
We focus on the invariant mass distribution of the $W$-boson pairs
as well as their total production cross section. For both the observable,
the choice of unphysical factorization and renormalization scales is taken to be
$\mu_R=\mu_F=\mu_0$. Here $\mu_0$ is the central scale which by default is chosen to be 
the invariant mass $Q$ of the $W$-boson pair.

The different input parameters that enter in this numerical study are taken as follows: 
The masses of the 
$W$-bosons is taken to be $ m_{w}=80.385$ GeV, while for its
couplings to other SM particles, we chose the Weinberg angle $\text{sin}^2\theta_\text{w} = (1 - m_w^2/\mz^2) = 0.222897$ and the
fine structure constant 
$\alpha \simeq 1/132.2332$ that
correspond to $G_F = 1.166379\times 10^{-5} \text{ GeV}^{-2}$.

We consider the four flavor scheme (4FS), hence a choice of the following PDFs with $n_f=4$ is taken:
NNPDF30\_lo\_as\_0118\_nf\_4 (for LO and LO+LL), NNPDF30\_nlo\_as\_0118\linebreak\_nf\_4 (for NLO and NLO+NLL) and NNPDF30\_nnlo\_as\_0118\_nf\_4 PDF (for NNLO and NNLO+NNLL) \cite{NNPDF:2017mvq}, all of them are taken from the {\tt LHAPDF} \cite{Buckley:2014ana}.
In the 4FS, the resonant top-quark contamination are avoided by removing bottom-quark emission subprocesses \cite{Gehrmann:2014fva}. Also, it has been noted that WW cross-sections for 4FS and 5FS agree within 1\%(2\%) for total production at 7(14) TeV LHC \cite{Gehrmann:2014fva,Grazzini:2017mhc}.

Except for the study of intrinsic PDF uncertainties, we stick to the central PDF set i.e. $iset=0$. The strong coupling constant at $\mz = 91.1876$ GeV is obtained from the respective PDF grids
and is then evolved to the desired scale, with the help of the routines taken from 
the {\tt LHAPDF}.


The NLO and the NNLO fixed order results are taken from the 
public package {\tt MATRIX} \cite{Grazzini:2017mhc}.
Moreover, at the NLO level these results are compared with those obtained from MadGraph \cite{Alwall_2011} and found that they agree well with each other. 
The numerical results from \texttt{MATRIX} at NNLO are verified by reproducing the results provided in the literature \cite{Gehrmann:2014fva,Grazzini:2017mhc}.
For the resummation, we have completely used in-house developed numerical codes with the corresponding two loop amplitudes taken from the {\tt VVamp} package \cite{Gehrmann:2015ora}.
%
To quantify the size of the higher order QCD corrections resulting from
both fixed order as well as the resummation, we define the following K-factors:
\begin{align}
& K_{\text{mn}}
=
	\frac{\sigma_{\text{N}^m\text{LO}}}{\sigma_{\text{N}^n\text{LO}}}
\,,~
R_{\text{mn}}
=
\frac{\sigma_{\text{N}^m\text{LO} + \text{N}^m\text{LL}}}{\sigma_{\text{N}^n\text{LO}}}~
,
\text{ and }
~
L_{\text{mn}}
=
\frac{\sigma_{\text{N}^m\text{LO} + \text{N}^m\text{LL}}}{\sigma_{\text{N}^n\text{LO}+\text{N}^n\text{LL}}}
~
        \label{eq:ratio}
\end{align}
We note that this definition of the ratios of the cross sections is in line with our previous analysis
carried out in \cite{Banerjee:2024xdh} for the case of on-shell $Z$-boson pair production.




\begin{figure}[ht!]
	\centerline{
		\includegraphics[scale =0.406]{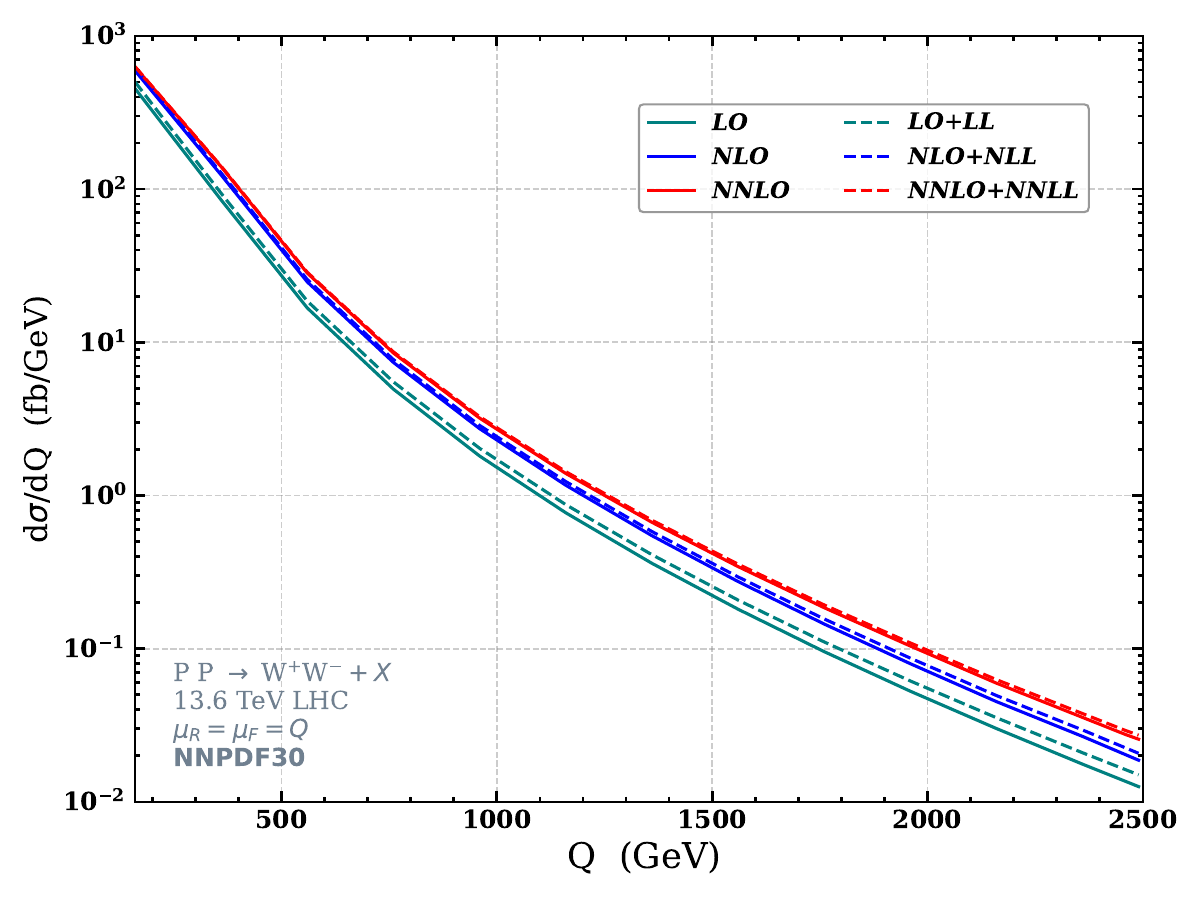}
		\includegraphics[scale =0.4]{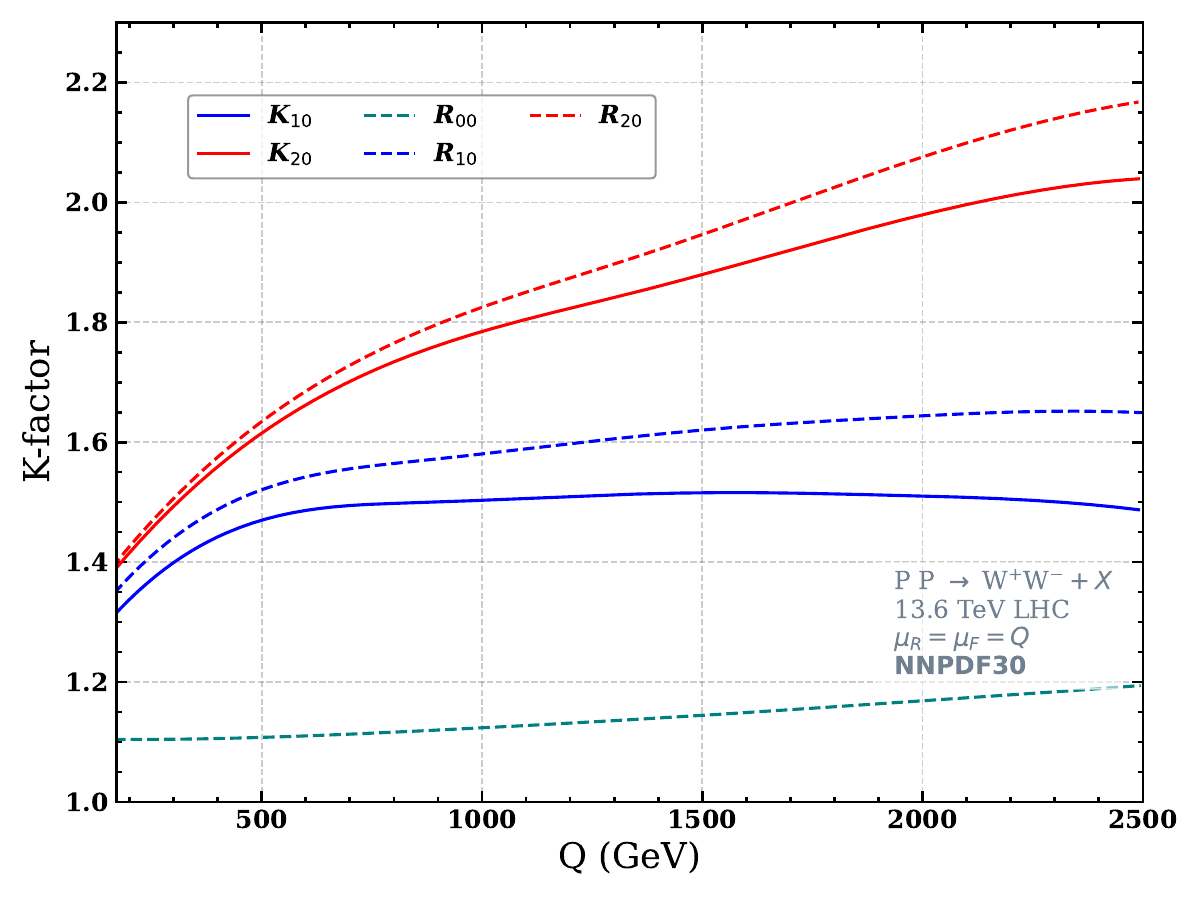}
	}
	\vspace{-2mm}
	\caption{\small{Resummed predictions for the invariant mass distribution (left) of the $W$-boson pair production and the corresponding K-factors(right) up to NNLO+NNLL.}}
	\label{fig:match_WW_inv}
\end{figure}

\begin{figure}[ht!]
	\centerline{
		\includegraphics[scale =0.5]{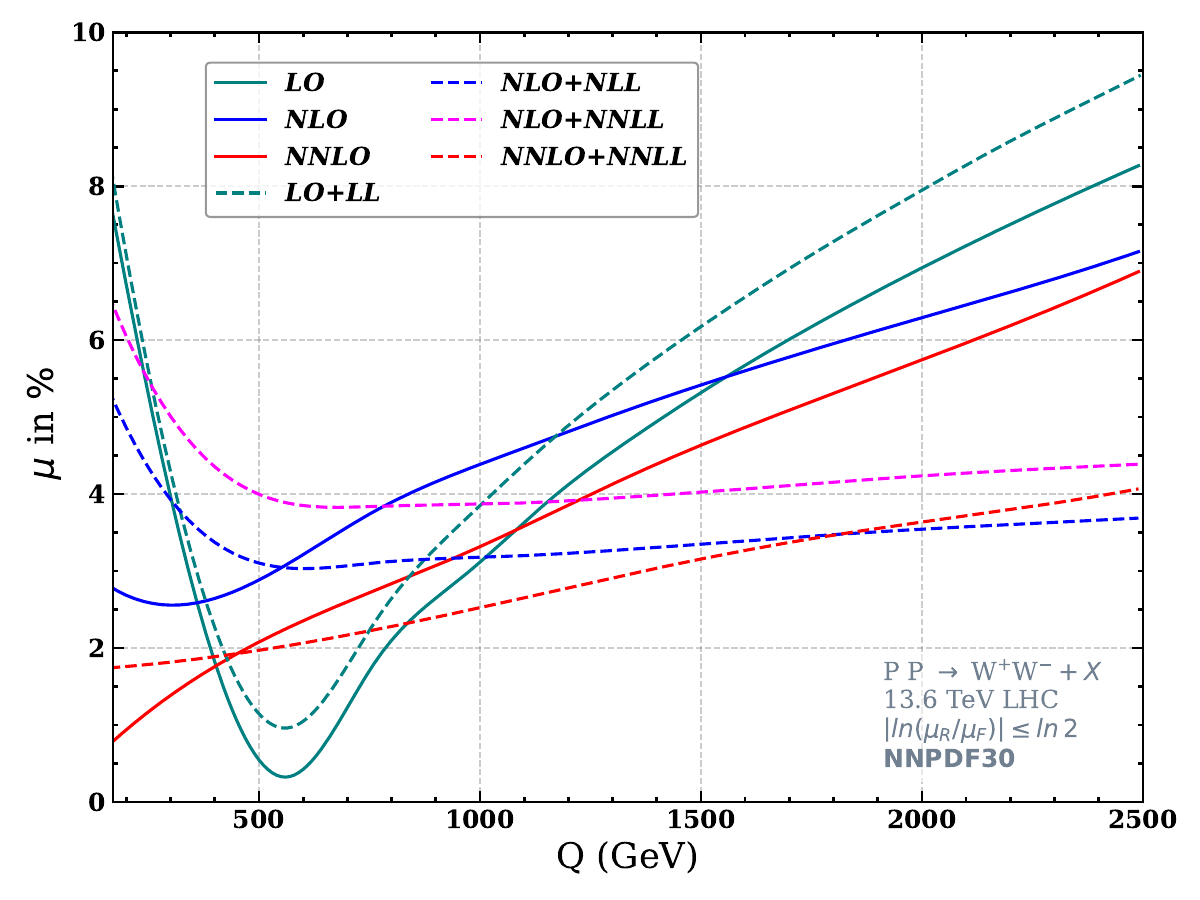}
	}
	\vspace{-2mm}
	\caption{\small{Seven point scale uncertainties for $W$-boson pair production up to NNLO+NNLL.}}
	\label{fig:match_WW_mu}
\end{figure}

\begin{figure}[ht!]
	\centerline{
		\includegraphics[scale =0.4]{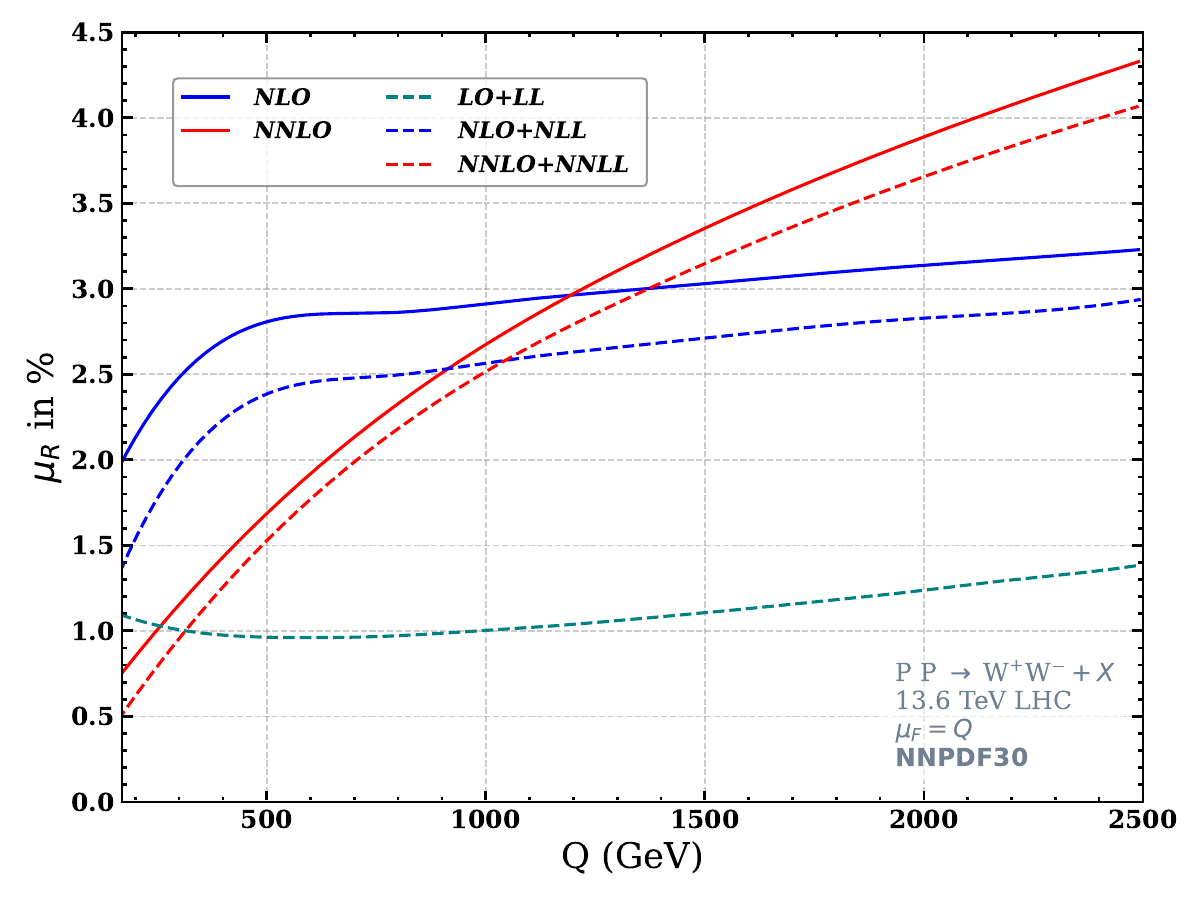}
		\includegraphics[scale =0.4]{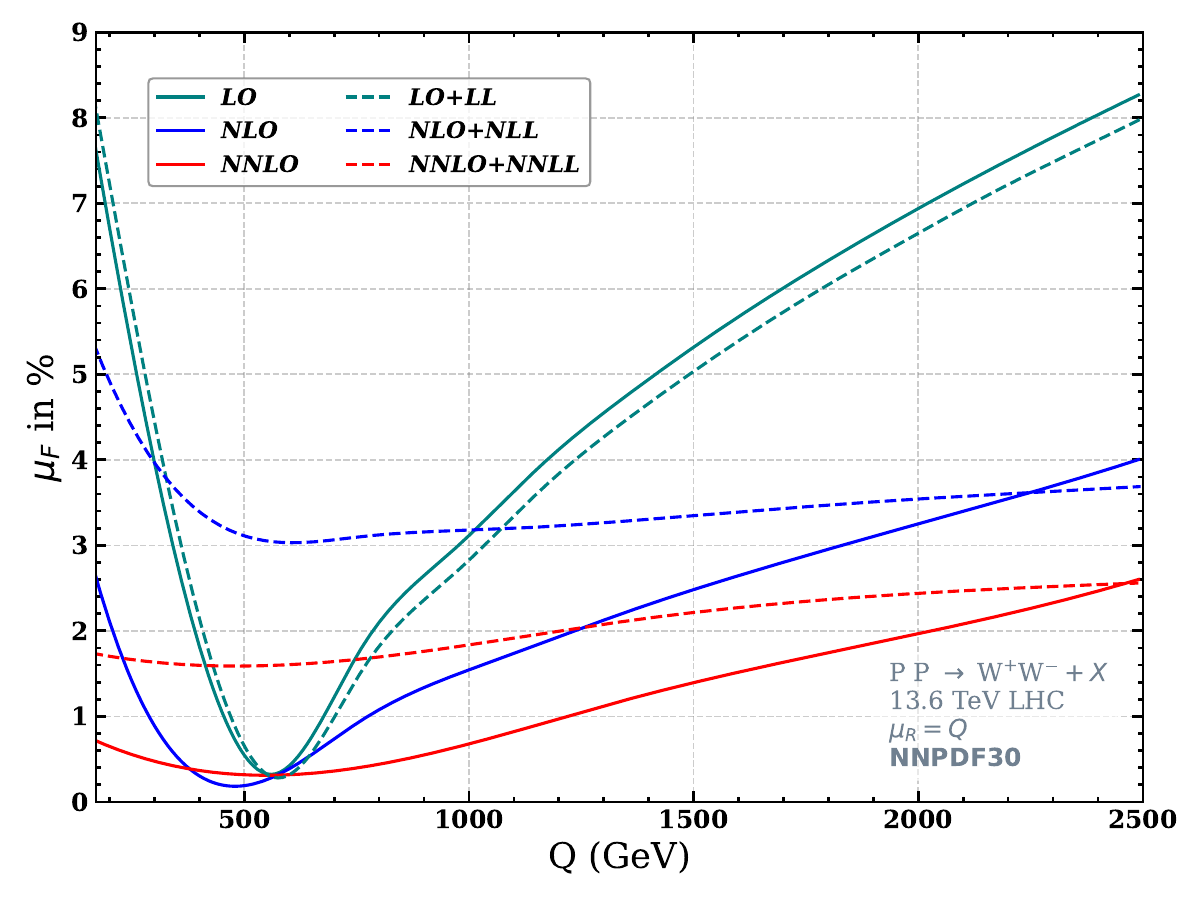}
	}
	\vspace{-2mm}
	\caption{\small{Renormalization(left) and factorization(right) scale uncertainties for $W$-boson pair production up to NNLO+NNLL.}}
	\label{fig:match_WW_mur}
\end{figure}

    

%

In \fig{fig:match_WW_inv}, we present the invariant mass distribution from the $2 m_W$  threshold  to $2500$ GeV for various orders in QCD up to NNLO$+$NNLL.
Due to the underlying parton fluxes that decrease with $Q$, the cross-sections rapidly fall by about four orders of magnitude for the kinematic region considered here.
The corresponding K-factors, defined above, are presented in the right panel.
In the rest of the discussion, for convenience, we quote all the enhancements due to both fixed order as well as resummed results in terms of the percentage defined with respect to LO. 
From $K_{10}$, we notice that the NLO order corrections enhance the LO predictions
by about $31\%$ at $Q=2 m_W$ and by about $49\%$ at $Q=2500$ GeV, with their contribution
remaining almost constant between $500$ GeV and $2500$ GeV.
The LL resummation enhances the LO results by as much as $19\%$ at $Q=2500$ GeV,
while the NLO+NLL resummation contributes about $64\%$ for the same $Q$ value.
In contrast, the NNLO corrections vary slowly but continuously and contribute 
about $38\%$ at $Q=2m_W$ to about $103\%$ at $Q=2500$ GeV, as can be seen 
from the behaviour of $K_{20}$. This obviously indicates the presence of additional 
subprocesses viz. $q_1 \bar{q_1} \to WW q_2 \bar{q_2}$ that enter the calculation
from NNLO onwards, with their contributions increasing with $Q$. 
This kind of behavior is known for $VV$ production process at hadron colliders as is reported in the literature \cite{Grazzini:2019jkl}. The inclusion of NNLL resummation makes the NNLO+NNLL corrections as big as $116\%$ at $Q=2500$ GeV.


Next, we quantify different sources of uncertainties in our theoretical predictions. The conventional 7-point scale uncertainties are estimated by varying the unknown renormalization $\mu_R$ and factorization $\mu_F$ scales around the central scale choice $\mu_0$ in the range from $\mu_0/2$ to $2 \mu_0$. These scale uncertainties are used to estimate the size of missing higher order corrections. The symmetric uncertainty is quantified as the maximum deviation of the cross sections from those obtained with the central scale choice. As mentioned previously, we take the central scale by default to be $\mu_0=Q$.

In \fig{fig:match_WW_mu}, we present these symmetric uncertainties in the invariant mass
distribution up to NNLO+NNLL in QCD. The uncertainty at LO is maximum of about $7.9\%$  at $Q=2 m_{w}$ and steeply falls to a minimum at around $Q=550$ GeV and then starts increasing with $Q$. On the contrary the scale uncertainties at both NLO and NNLO level  keep increasing monotonically with $Q$. 
Moreover, the scale uncertainties at NNLO are smaller than those at NLO for the entire range of $Q$ considered here, as can be seen from Fig.~\ref{fig:match_WW_mu}.
The uncertainty in NLO (NNLO) at $Q=2 m_{w}$ is
around $2.8\%$ ($0.7\%$) while that at $Q=2500$ GeV is $7.1\%$ ($6.9\%$).

In the case of resummed results presented  in the Fig.~[\ref{fig:match_WW_mu}], specifically in the low $Q$-region where the process-dependent regular terms as well as other parton channels give sizable corrections (and together with the fact that they can not be resummed), the 
resummation does not improve the scale uncertainties. 
However, as $Q$ increases, in the present case for $Q > 560$ GeV, the resummation of large threshold logarithms reduces these scale uncertainties.
While, for NLO(NNLO), the scale uncertainties reach up to  $ 7.1\% (6.8)\%$ at $Q=2500$ GeV, the corresponding ones at NLO+NLL(NNLO+NNLL) level get reduced to $3.7 \% (4.1\%)$. 
It is also to be noted that the NLO+NNLL results for this process have been available for quite some time now \cite{Dawson:2013lya}. 
However, the scale uncertainties in these NLO+NNLL results are larger than those in the NLO+NLL case for the entire range of $Q$ considered here, as can be seen from the Fig.~[\ref{fig:match_WW_mu}]. 
This also indicates the requirement for the inclusion of the two loop process dependent contributions, via $g_{02}$, in the resummation, followed by a systematic matching to the relevant fixed order NNLO results. 
After performing this, the $7$-point scale uncertainties at NLO+NNLL level significantly get reduced from about $6.6\% (4.0\%)$ at $Q= 2m_w $ (550 GeV) to about $1.7\%$($2.0\%$) at NNLO+NNLL level.

In the \fig{fig:match_WW_mur}, we separately study the uncertainties in the cross-sections by varying one of these scales in the range [$\mu_0$/2, 2$\mu_0$], while keeping the other fixed at $\mu_0$. 
In the left (right) panel, we present the renormalization (factorization)  scale uncertainties in the invariant mass distribution up to NNLO+NNLL.
Unlike the case of LO where there is no dependence on $\mu_R$, the renormalization scale enters LL through the resummation of leading logarithms.
Consequently,  the LO+LL shows an enhanced scale uncertainties than LO, as is evident from the left panel of Fig.~[\ref{fig:match_WW_mur}]. Starting from NLO, the scale $\mu_R$ enters the fixed order results through $a_s(\mu_R)$. 
In going beyond LO+LL (for a fixed $\mu_F$), as a result of the resummation of threshold logarithms to higher logarithmic accuracy,  the scale uncertainties at NLO+NLL and NNLO+NNLL level will become smaller than the corresponding ones in the fixed order NLO and NNLO results, see \fig{fig:match_WW_mur}. 

In the right panel, we present the scale uncertainties in the invariant mass distribution by varying the factorization scale $\mu_F$ while keeping the other scale $\mu_R$ fixed at $\mu_0$.
This factorization scale enters both the PDFs as well as the parton level coefficient functions. 
Although the parton level coefficient functions include the threshold corrections at a given logarithmic accuracy to all orders in $a_s$, the undelrying PDFs that are used in the calculation are extracted at a given order in $a_s(\mu_R)$. 
Thus, the factorization scale uncertainty is not expected to decrease as a result of 
threshold resummation as is depicted in the right panel of \fig{fig:match_WW_mur},
see  \cite{
AH:2019phz,
AH:2020cok,
AH:2020iki,AH:2021kvg,AH:2021vdc,Bhattacharya:2021hae,Das:2022zie,Ravindran:2022aqr,Das:2024auk,Banerjee:2024xdh,Banerjee:2025ydi,Das:2025wbj,Bhattacharya:2025rqk,Banerjee:2025fsj}
for more details.

Although, the scale uncertainties are estimated by varying the $\mu_R$ and $\mu_F$ around
a specific central scale choice $\mu_0 = Q$, it will be interesting to study how these
theory uncertainties will change for a different choice of the central scale.
For this study, we take three different
choices for the central scale $\mu_0$, namely, $\mu_0 = Q/2, Q \text{ and } 2Q $, and present the $7$-point scale uncertainties for the invariant mass distribution in Tab.~[\ref{tab:tableWW_Q}]. We notice that the scale uncertainties in both NNLO as well as
NNLO+NNLL level for the central choice $\mu_0 = Q/2$ are larger than those for the case of $\mu_0=Q$, which in turn are found to be  larger than the case when $\mu_0 = 2Q$. 
This behaviour is observed for the entire $Q$ range considered in our analysis, for brevity we present the results for  some values of $Q$.
\begin{table}[ht!]
\begin{center}

\begin{tabular}{|c|cc|cc|cc|}
\hline
\multirow{2}{*}{Q(GeV)} & \multicolumn{2}{c|}{$\mu_{0}$ = $2$Q}                               & \multicolumn{2}{c|}{$\mu_{0}$ = Q}                                & \multicolumn{2}{c|}{$\mu_{0}$ = Q/$2$} \\  \cline{2-7}
                        & \multicolumn{1}{c|}{${\text{NNLO}}$} & ${\text{NNLO+NNLL}}$          & \multicolumn{1}{c|}{${\text{NNLO}}$} & ${\text{NNLO+NNLL}}$           & \multicolumn{1}{c|}{${\text{NNLO}}$}   & ${\text{NNLO+NNLL}}$  \\ \hline
165                     & \multicolumn{1}{c|}{0.71} & 1.37                                     & \multicolumn{1}{c|}{0.74} & 1.74                                                 & \multicolumn{1}{c|}{0.99}   & 2.14       \\ \hline
565                     & \multicolumn{1}{c|}{2.15} & 2.03                                     & \multicolumn{1}{c|}{2.25} & 2.02                                                 & \multicolumn{1}{c|}{2.50}   & 2.23       \\ \hline
1365                    & \multicolumn{1}{c|}{3.97} & 2.80                                     & \multicolumn{1}{c|}{4.29} & 2.99                                                 & \multicolumn{1}{c|}{4.49}   & 3.28       \\ \hline
1765                    & \multicolumn{1}{c|}{4.76} & 3.19                                     & \multicolumn{1}{c|}{5.22} & 3.43                                                 & \multicolumn{1}{c|}{5.48}   & 3.72       \\ \hline
\end{tabular}
\caption{The 7-point scale uncertainty (in $\%$) for the different central scale choices at different Q values.}
\label{tab:tableWW_Q}
\end{center}
\end{table}


%

Apart from the theoretical uncertainties
due to the unphysical scales in the calculation, there will also be uncertainties due to the non-perturbative 
PDFs.  Here, using the {\tt LHAPDF} routines, we estimate the intrinsic PDF uncertainties both in the fixed
order NNLO results as well as the resummed NNLO+NNLL predictions for the invariant mass distribution, and present the same 
for different $Q$ values in Tab. [\ref{tab:tableWW_pdf}]. The PDFs are well constrained in the low $x$-region (where $x$
is the Bjorken variable) compared to the high $x$-region, consequently the PDF
uncertainties are expected to increase with $Q$. These uncertainties are about $1.52\%$ at $Q=165$ GeV and are found to increase to about $2.78\%$ around $2000$ GeV.
\begin{table}[ht!]
\begin{center}
\begin{tabular}{|l|l|l|l|} \hline
Q (GeV)        & 165 & 1365 & 1965  \\ \hline
sym err (NNLO) \% &  1.52  &  2.33   & 2.80 \\ \hline
sym err (NNLO$+$NNLL) \% &  1.52  & 2.32   & 2.78 \\ \hline
\end{tabular}
\caption{Intrinsic PDF error of NNPDF30\_nnlo\_as\_01180\_nf\_4 PDFs.}
\label{tab:tableWW_pdf}
\end{center}
\end{table}

We also study the total cross-section for the production of $W$-boson pairs at the LHC by integrating the invariant mass distribution from $Q_{min}=2m_w$ to $Q_{max}=\sqrt{S}$.  
For LHC energies, the total production cross-sections for the process under consideration are quite large. Specifically they are larger than those for the $Z$-boson pair production process due to the underlying parton fluxes, the corresponding couplings of the quarks to the gauge bosons and due to the presence of additional $s$-channel contributions that are absent in the case of $Z$-boson pair production. 
For the current LHC energies, the quark initiated subprocess even at LO itself is about seven times larger for the $WW$ case than for $ZZ$ case \cite{Banerjee:2024xdh}. 


\begin{table}[ht!]
\begin{center}
{\scriptsize		
\resizebox{15.0cm}{2.3cm}{
\begin{tabular}{|l|l|l|l|l|l|l|}
\hline
$\sqrt{S}$ (TeV)                     & $7.0$ TeV                            & $8.0$ TeV                            & $13.0$ TeV                            & $13.6$ TeV                          & $14.0$ TeV                            & $100.0$ TeV                             \\ \hline \hline
LO                                   & $31.116_{-1.69\%}^{+1.13\%}$ & $37.584_{-2.39\%}^{+1.78\%}$ & $71.825_{-4.86\%}^{+4.13\%}$  & $76.086_{-5.08\%}^{+4.35\%}$  & $78.941_{-5.23\%}^{+4.49\%}$  & $747.343_{-13.96\%}^{+13.82\%}$ \\ \hline
NLO                                  & $42.309_{-2.02\%}^{+2.45\%}$ & $51.410_{-2.02\%}^{+2.45\%}$ & $100.357_{-2.04\%}^{+2.47\%}$ & $106.518_{-2.04\%}^{+2.48\%}$ & $110.667_{-2.05\%}^{+2.48\%}$ & $1130.054_{-5.63\%}^{+3.84\%}$  \\ \hline
$\text{NNLO}_{q\bar{q}}$             & $46.286_{-1.36\%}^{+1.37\%}$ & $ 56.201 ^{ + 1.33 \% }_{ -1.32 \% }$ & $ 110.487 ^{ + 1.35 \% }_{ -1.34 \% }$ & $ 117.151 ^{ + 1.35 \% }_{ -1.33 \% }$ & $ 122.008 ^{ + 1.40 \% }_{ -1.34 \% }$ &  $ 1248.098 ^{ + 1.69 \% }_{ -1.81 \% }$ \\ \hline
NNLO                                 & $ 47.137 ^{ +1.81 \% }_{ -1.63 \% }$ & $ 57.340 ^{ +1.75 \% }_{ -1.62 \% }$ & $ 113.523 ^{ +1.91 \% }_{ -1.75 \% }$ & $ 120.459 ^{ +1.93 \% }_{ -1.76 \% }$ & $ 125.500 ^{ +1.98 \% }_{ -1.77 \% }$ & $ 1323.434 ^{ +2.84 \% }_{ -2.47 \% }$  \\ \hline
LO+LL                                & $ 34.710 ^{ +1.44 \% }_{ -2.02 \% }$ & $ 41.824 ^{ +2.08 \% }_{ -2.71 \% }$ & $  79.387 ^{ +4.43 \% }_{ -5.16 \% }$ & $  84.055 ^{ +4.65 \% }_{ -5.39 \% }$ & $  87.183 ^{ +4.79 \% }_{ -5.53 \% }$ & $   814.504 ^{ +14.12 \% }_{ -14.20 \% }$  \\ \hline
NLO+NLL                              & $ 43.766 ^{ +3.13 \% }_{ -3.18 \% }$ & $ 53.134 ^{ +3.28 \% }_{ -3.38 \% }$ & $ 103.464 ^{ +3.81 \% }_{ -4.14 \% }$ & $ 109.794 ^{ +3.86 \% }_{ -4.21 \% }$ & $ 114.057 ^{ +3.90 \% }_{ -4.26 \% }$ & $ 1158.204 ^{ +6.38 \% }_{ -7.80 \% }$  \\ \hline
$\text{NNLO}_{q\bar{q}}+\text{NNLL}$ & $ 46.684 ^{ +1.20 \% }_{ -1.21 \% }$ & $ 56.762 ^{ +1.00 \% }_{ -1.43 \% }$ & $ 111.341 ^{ +1.19 \% }_{ -1.48 \% }$ & $ 118.059 ^{ +1.19 \% }_{ -1.51 \% }$ & $ 122.947 ^{ +1.24 \% }_{ -1.52 \% }$ & $ 1255.799 ^{ +1.76 \% }_{ -2.72 \% }$  \\ \hline
NNLO+NNLL                            & $ 47.536 ^{ +1.58 \% }_{ -1.45 \% }$ & $ 57.185 ^{ +1.57 \% }_{ -1.45 \% }$ & $ 114.378 ^{ +1.75 \% }_{ -1.58 \% }$ & $ 121.367 ^{ +1.77 \% }_{ -1.60 \% }$ & $ 126.439 ^{ +1.82 \% }_{ -1.61 \% }$ & $ 1331.135 ^{ +2.72 \% }_{ -3.28 \% }$
  \\ \hline
 \hline
CMS \{13.0 TeV ($35.9$ fb$^{-1}$)\}~\cite{CMS:2020mxy} & \multicolumn{6}{c|}{$117.6 \pm 6.8$ pb }\\ \hline 
ATLAS \{13.0 TeV ($140$ fb$^{-1}$)\}~\cite{ATLAS:2025dhf} & \multicolumn{6}{c|}{ $ 127.7 \pm 4$ pb }\\ \hline 
\end{tabular}
}
\caption{\small{Inclusive cross section (in pb) for $W^{+}W^{-}$ boson pair production for different center of mass energies of the incoming protons, along with the corresponding 7-point scale uncertainties.. The measurements form ATLAS and CMS at 13 TeV are also given. The total uncertainty accounts for the statistical, systematic and luminosity uncertainties involved in the measurement.}}
\label{tab:tableWW} }
\end{center} 
\end{table}

The results are presented in the \tab{tab:tableWW} for different
$\sqrt{S}$ values, along with the corresponding scale uncertainties. For this observable
of total cross section, bulk of the contribution comes from the low $Q$-region where
the parton fluxes are high.  As discussed before, the scale uncertainties in the low 
$Q$-region don't get improvement because of the resummation, consequently the scale uncertainties for the total production cross section are larger in the resummation case than in the fixed order case. It is worth mentioning that the gluon fusion channel
contributes to this process starting from NNLO, and because of the large gluon fluxes
for the current LHC energies, its contribution to the total cross section 
at this level of precision is important.
We present the total cross section at the two loop level in both ways i.e. with this gluon fusion contribution (NNLO) and without (NNLO$_{q\bar{q}}$).
The scale uncertainty at LO is about $5.1\%$ and it gets reduced to about $1.35\%$ at
NNLO$_{q\bar{q}}$ for $13.6$ TeV LHC as can be seen from the  \tab{tab:tableWW}. 
On the contrary, for the resummation case, the uncertainties about $5.4\%$ at LO+LL
get reduced to $1.5\%$ at NNLO$_{q\bar{q}}$+NNLL.
This gluon fusion channel contributes about $4.5\%$ of the LO quark initiated process and is about $3.3$ pb for $13.6$ TeV LHC energy. 
Because of this gluon fusion channel,
there will be additional source of contribution to the scale uncertainties 
from NNLO on wards.
Upon the inclusion of gluon fusion channel the scale uncertainties at NNLO+NNLL are found to get to increased to $1.61\%$. However, with increasing $Q_{\rm min}$ in the total production cross-section, the threshold logarithms start dominating the cross-sections and tend to reduce the scale uncertainties in the NNLO+NNLL results, as evident from \fig{fig:match_WW_mu}. 
Further, the one-loop corrections for the gluon fusion channel, in the massless quark limit, are already available and are about $60\%$ for $13$ TeV LHC energy 
as reported in \cite{Caola:2016trd}.

Apart from this gluon fusion channel, there will be photon fusion contribution to this $W$-boson pair production process. 
The $W$-bosons being charged particles, unlike the $Z$-bosons, the photons directly couple at the tree level itself. For completeness, we report on these contributions below. For the $13.0$ TeV LHC, the photon fusion contribution is about $0.827$ pb \cite{Alwall_2011}.
The photon distributions inside the proton are available from \cite{Manohar:2016nzj,Bertone:2017bme}.
At this precision level, the NLO electroweak corrections are also important and 
are available now for both $q \bar{q}$ initiated as well as photon fusion subprocesses (including the $q\gamma$ contributions). The net contribution of these NLO electroweak corrections is found to be around $-2\%$ of the LO \cite{Grazzini:2019jkl}.


Finally, it is worth noting that both ATLAS and CMS experiments have measured the total production cross sections, and the reported central values differ from each other. With the integrated luminosity of about $35.9 \text{ fb}^{-1}$, the CMS measurement is about $117.6\pm6.8$ pb \cite{CMS:2020mxy}, while ATLAS reported a value of about $137.9\pm 10.0$ pb \cite{ATLAS:2019rob} for $13$ TeV LHC. In the recent measurement by ATLAS with increased integrated luminosity of about  $140 \text{ fb}^{-1}$, the measurement stands at $127.7\pm 4$ pb \cite{ATLAS:2025dhf}, which agrees with CMS ($35.9 \text{ fb}^{-1}$)\cite{CMS:2020mxy} reported cross-section within errors.



\begin{figure}[ht!]
   \centering
   \begin{minipage}{0.46\textwidth}
        \centering
        \includegraphics[scale=0.35]{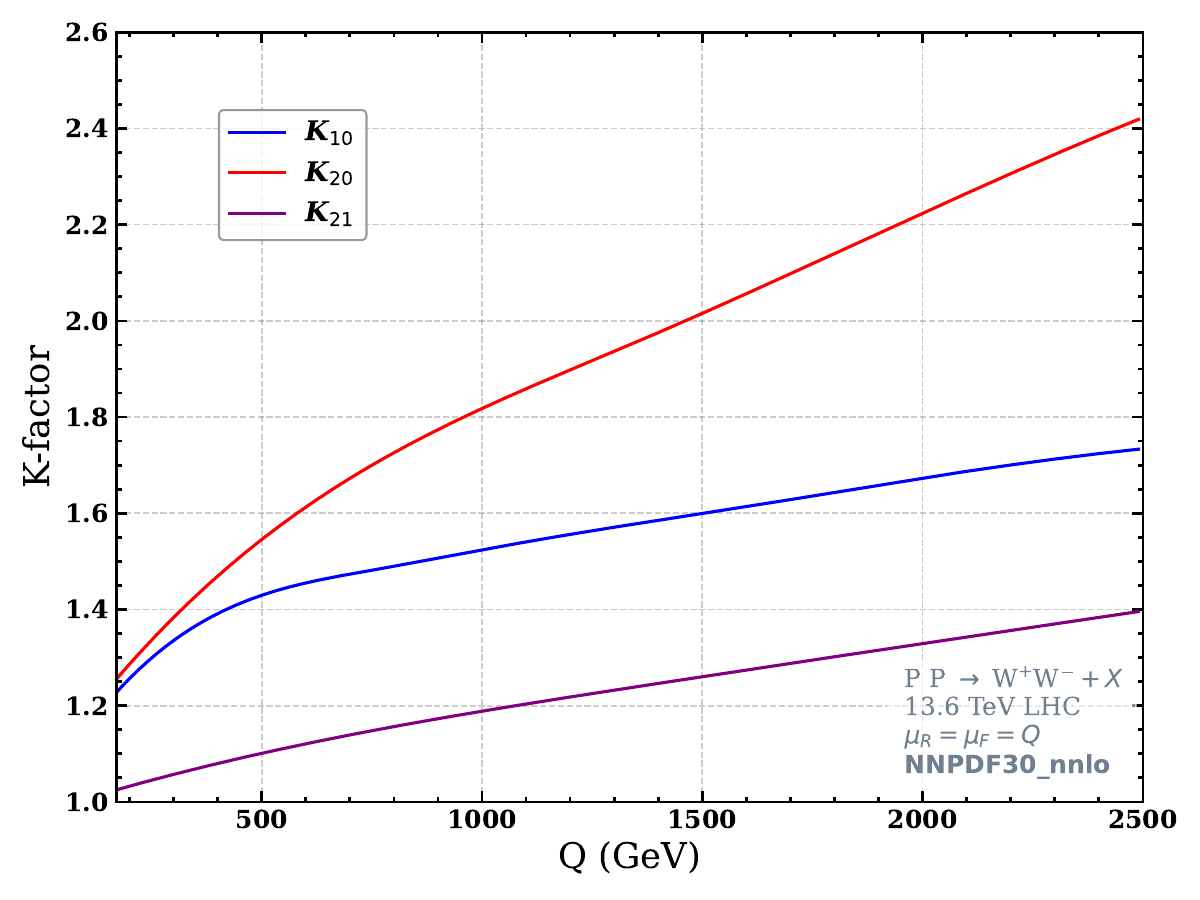}
    \end{minipage}
    \hfill
    \begin{minipage}{0.46\textwidth}
        \centering
        \includegraphics[scale=0.35]{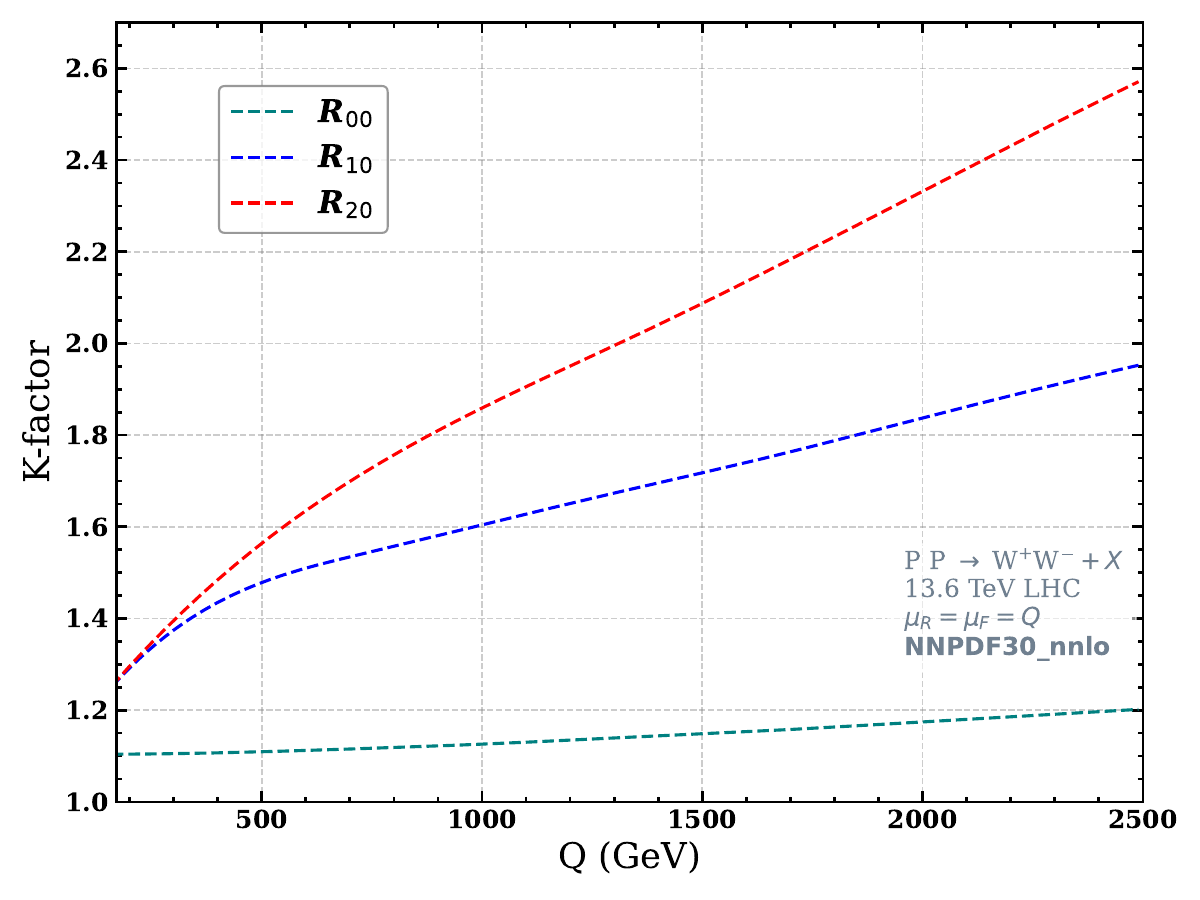}
    \end{minipage}
    
    \begin{minipage}{0.46\textwidth}
        \centering
        \includegraphics[scale=0.35]{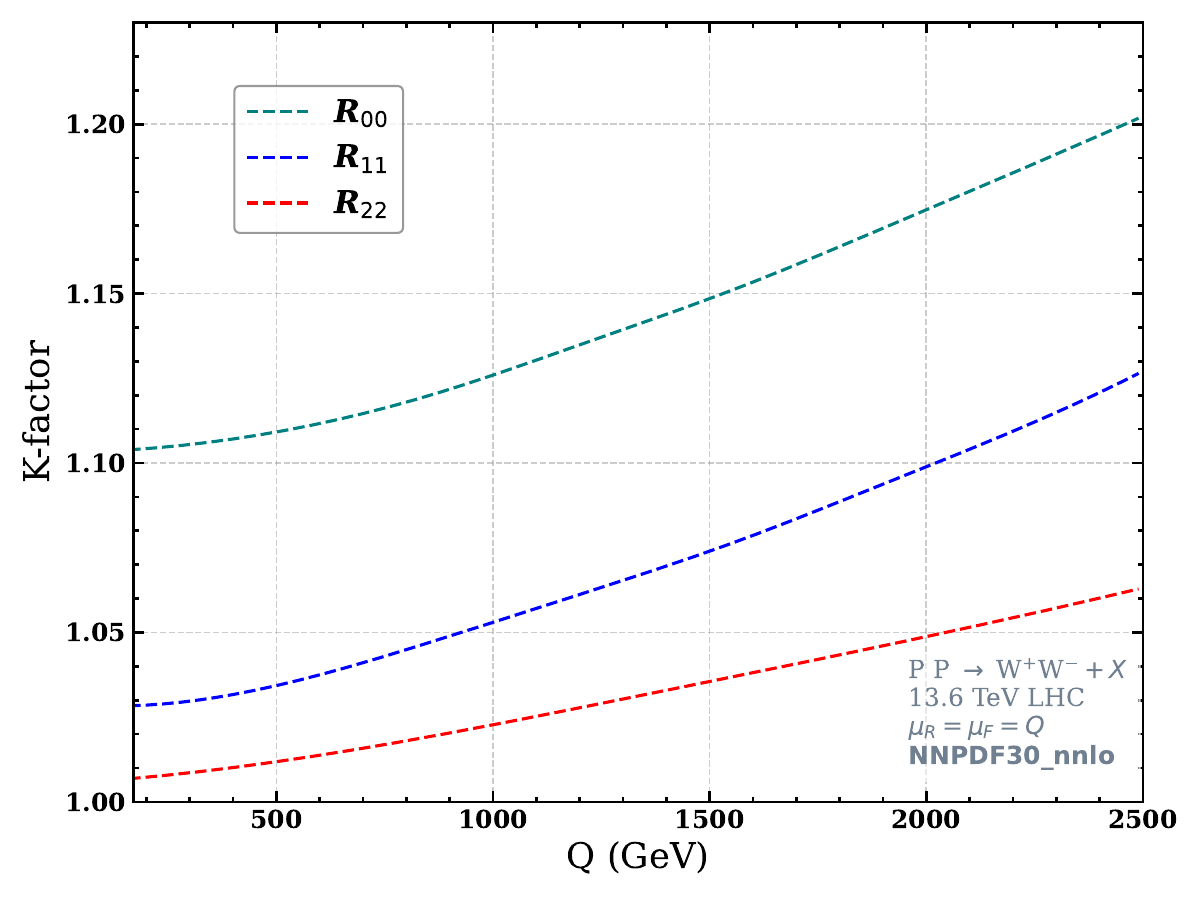}
    \end{minipage}
    \hfill
    \begin{minipage}{0.46\textwidth}
        \centering
        \includegraphics[scale=0.35]{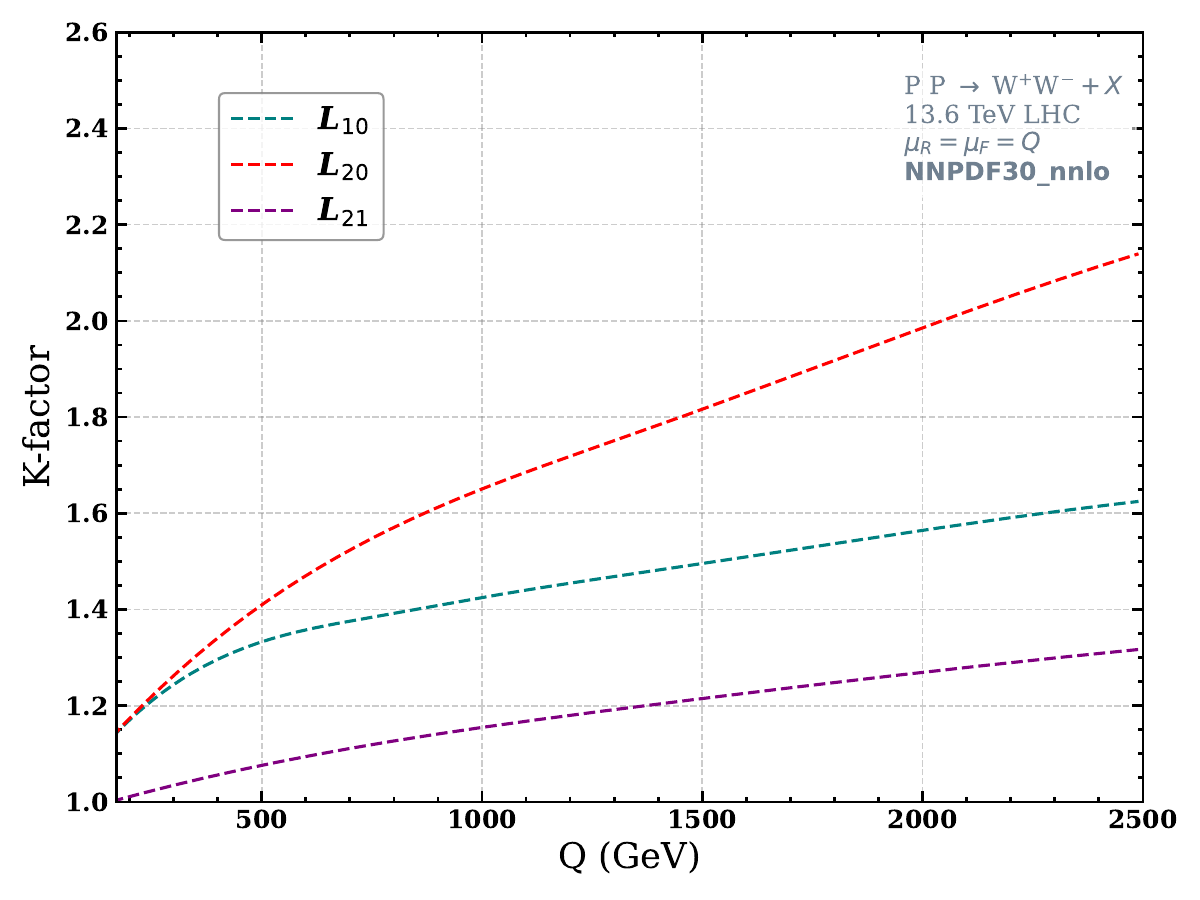}
    \end{minipage}
       \vspace{-2mm}
	\caption{\small{K-factors as defined in \eq{eq:ratio} but using same NNLO PDFs at various orders in the perturbation theory.}}
\label{fig:match_WW_kfacnnlo}
\end{figure}

Further, in order to study the rate of convergence of the perturbative coefficient functions both in the fixed order as well as in the resummed predictions, it is useful to fix the PDF sets and study the K-factors like 
$K_{i,j}$, $R_{i,j}$ and $L_{i,j}$ (defined in \eq{eq:ratio}), 
which are quite informative. 
Here, we chose the NNLO PDF sets, along with the respective $\as$ from {\tt LHAPDF}, and present the results up to NNLO+NNLL level in the \fig{fig:match_WW_kfacnnlo}.
In the top left panel, we present the fixed order K-factors, $K_{ij}$.
For $i > j$, these ratios $K_{ij}$ give the enhancement of the higher order
corrections over and above any given $j^{\rm th}$ fixed order result, and give
us the information about how fast the fixed order results are converging with 
increasing $j$.
From the plot, one can see that the first order corrections
can be as large as $73\%$ of LO for $Q=2500$ GeV, while the second order corrections 
contribute a maximum of about $40\%$ of NLO result.
Moreover, each of these $K_{ij}$ is found to increase slowly with $Q$.
In the top right panel of \fig{fig:match_WW_kfacnnlo}, we present the resummed K-factors $R_{i0}$ that are defined with respect to the LO predictions.
Unlike the $K_{i0}$, these ratios give the size of both fixed order as well as
the threshold logarithmic contributions.
In the bottom left panel of \fig{fig:match_WW_kfacnnlo}, we plot $R_{ii}$ that 
simply estimate the size of the threshold logarithms over a given fixed order result. By definition, they are supposed to decrease with increasing $i$ as depicted in the figure. However, for any given $i$, these ratios are observed to slowly increase with $Q$, which in the present case indicate that the NNLL resummation contributes 
an additional $6.2\%$ of NNLO at $Q=2500$ GeV.
Finally, in the bottom right panel, we present the ratios  $L_{ij}$. Similar to $K_{ij}$, these ratios give the information about the
contribution of higher order corrections but also estimates the impact of further sub-leading threshold logarithms over and above a given logarithmic accuracy through resummation and matching to the appropriate fixed order results. 
Here, the denominator already captures the threshold logarithms through resummation to
a given logarithmic accuracy, and hence these ratios are expected to be smaller than $K_{ij}$.

\section{Summary}\label{sec:conclusion}
In our article, we have presented threshold resummation for the production of a pair of on-shell $W$-bosons. We have performed the resummation at NNLO+NNLL accuracy, and give our results up to $Q=2500$ GeV. 
On the top of the available NNLO results, the contribution of NNLL resummed results is found to be about $1\%$ of LO  at $Q=165$ GeV and rises to as large as $13\%$ at $Q=2500$ GeV.
We have further compared and contrasted these NNLO+NNLL resummed results with those at NLO+NNLL accuracy,
and find that the former have less scale uncertainties throughout the invariant mass region considered here.

In addition, we have investigated the effect of the intrinsic PDF uncertainty. While the PDF uncertainty at NNLO is 2.80\% at $Q=1965$ GeV, the corresponding one at NNLO+NNLL is about 2.78\%. In contrast, for the invariant mass distribution, 7-point scale uncertainties decrease from 5.7\% at NNLO to 3.6\% at NNLO+NNLL, at the same value of Q. This highlights the importance of our resummation results.
We have also studied the convergence of the perturbative series by evaluating the fixed order and resummed results using the same NNLO order PDFs. Moreover, we investigate the behaviour of the seven point scale uncertainties by changing the choice of central scale from $Q/2$ to $2Q$, and observe that the choice of central scale $Q$ has smaller uncertainties both for fixed-order as well as resummed results. The present results at NNLO+NNLL are the most precise results available to-date in the perturbative QCD for the invariant mass distribution of $W$-boson pair production process and are expected to augment the precision studies at the current LHC as well as at the future hadron colliders.

\section*{Acknowledgements}
 P.B. acknowledges A. Papa for useful discussion.
 The research work of M.C.K. is supported by SERB Core 
Research Grant (CRG) under the project CRG/2021/005270.
The work of P.B. is supported by the INFN/QFT@COLLIDERS project (Italy). 
We acknowledge National Supercomputing
Mission (NSM) for providing computing resources of ‘{\tt PARAM Kamrupa}’ at IIT Guwahati, which is implemented by C-DAC and supported by the Ministry of Electronics and Information Technology (MeitY) and Department of Science and Technology (DST), Government of India,
where most of the computational work has been carried out. P.B. acknowledges the Department of Physics at the University of Calabria for providing computing resources. 

\pagebreak
\appendix
\section{Resummation coefficients}
\label{appendixa}
The process-dependent $g_0$ coefficients defined in \eq{eq:g0} are given as 
(defining $L_{qr} = \ln \left(Q^2/\mur^2 \right), L_{fr} = \ln \left( \muf^2/\mur^2\right)$),

\begin{align}\label{eq:g0}
	\begin{autobreak} 
		g_{01} = 
		2 \mathcal{A}_{(0,1)} 
		+ 2  \Cf \bigg\{ 
		- 3 \Lfr 
		+ \Lqr^2
		+ \z2
		\bigg\},
	\end{autobreak} 
	\\ 
	\begin{autobreak} 
		g_{02} = 
		\mathcal{A}_{(1,1)} 
		+ 2 \mathcal{A}_{(0,2)}
		+ \Cf \nf \bigg\{ 		
		- \frac{328}{21}
		+ \frac{2}{3}\Lfr
		+ \frac{112}{27}\Lqr
		- \frac{20}{9}\Lqr^2
		- \frac{10}{9}\z{2}
		+ \frac{16}{3}\Lfr \z{2}
		- \frac{4}{3}\Lqr \z{2}
		+ \frac{32}{3}\z{3}
		\bigg\}
		+ \Cf^2 \bigg\{ 
		- 3 \Lfr
		+ 18 \Lfr^2
		- 12 \Lfr \Lqr^2
		+ 2 \Lqr^4 
		+ 12 \Lfr \z{2}
		+ 4 \Lqr^2 \z{2}
		+ 2 \z{2}^2
		- 48 \Lfr \z{3}
		\bigg\}
		+ \Cf \bigg\{
		+ 3 \beta_{0} \Lfr^2
		- \frac{2}{3} \beta_{0} \Lqr^3
		-12 \Lfr \mathcal{A}_{(0,1)}
		+ 4 \Lqr^2 \mathcal{A}_{(0,1)}
		- 2 \beta_{0} \Lqr \z{2}
		+ 4 \mathcal{A}_{(0,1)} \z{2}
		+ \frac{46}{3}\beta_{0} \z{3}
		\bigg\}
		+\Cf \Ca \biggl\{
		+ \frac{2428}{81}
		- \frac{17}{3} \Lfr
		-\frac{808}{27} \Lqr
		+ \frac{134}{9} \Lqr^2
		+ \frac{67}{9} \z{2}
		- \frac{88}{3}\Lfr \z{2}
		+ \frac{22}{3} \Lqr \z{2}
		- 4 \Lqr^2 \z{2}
		- 12 \z{2}^2
		-\frac{176}{3} \z{3}
		+ 24 \Lfr \z{3}
		+ 28 \Lqr \z{3}
		\biggr\}
	\end{autobreak} 
\end{align}

Here, we define
\begin{align}
    \begin{autobreak}
	    \mathcal{A}_{(m,n)} = \frac{\mathcal{M}_{(m,n)}^{\text{fin}}}{\mathcal{M}_{(0,0)}}
    \end{autobreak}
\end{align}
where,
\begin{align}
    \begin{autobreak}
	    \mathcal{M}_{(m,n)}^{\text{fin}} \equiv \langle \mathcal{M}_{m}  \rvert \mathcal{M}_{n} \rangle
    \end{autobreak}
\end{align}
and $ \rvert \mathcal{M}_{n} \rangle$ represents the UV-renormalized, IR-finite virtual amplitude at the n-th order in $\as$, as given in Eq. (C3) of Ref. \cite{Ahmed:2020nci}.
The one loop virtual contribution for $W^{+}W^{-}$ pair production ($u$-type) is ,

\begin{align}
    \mathcal{M}_{0,1} = \frac{\alpha_{s}}{4 \pi} B_{f} N C_{f} \left( c_{tt} \mathcal{F}_{1} - c_{ts}\mathcal{J}_{1} + c_{ss}\mathcal{K}_{1} \right)    
\end{align}
\begin{align}
    \mathcal{F}_{1} = \mathcal{I}_{1} \mathcal{F}_{0} + \mathcal{F}_{1}^{fin} , \quad
    \mathcal{J}_{1} = \mathcal{I}_{1} \mathcal{J}_{0} + \mathcal{J}_{1}^{fin} , \quad 
    \mathcal{K}_{1} = \mathcal{I}_{1} \mathcal{K}_{0} + \mathcal{K}_{1}^{fin}
\end{align}
where,
\begin{align}
    \mathcal{I}_{1} = -\frac{e^{\epsilon \gamma_{E}}}{\Gamma(1-\epsilon)}
                        \left( \frac{\mu^2}{s} \right)^{\epsilon}\left\{\frac{1}{\epsilon^2} + \frac{1}{\epsilon}\left(i \pi -\frac{3}{2} \right) \right\}\Cf .
\end{align}

\begin{align}
    \mathcal{F}_{1}^{fin} &  = \frac{1}{((-1 + 
     x)^5 x (1 + x) y^2)} \big[ 4 (-8 x^4 y (3 + 2 y^2) \zeta_{2} + 
     y (18 + y^2 (4 - 3 \zeta_{2}) - 12 \zeta_{2} \nonumber \\ &
     - 4 y (-4 + 3 \zeta_{2})) + x^8 y (-18 + 4 \zeta_{2} + 4 y (-4 + 3 \zeta_{2}) + y^2 (-4 + 3 \zeta_{2})) + x^7 (18  \nonumber \\ &
    + y^2 (50 - 31 \zeta_{2}) + y^4 (4 - 3 \zeta_{2}) - 4 \zeta_{2} - 4 y^3 (-4 + 3 \zeta_{2}) - 8 y (-9 + 4 \zeta_{2})) + 2 x^6  \nonumber \\ & 
    (-36 + 8 \zeta_{2} + 6 y^2 (-4 + 3 \zeta_{2}) + 4 y (-13 + 5 \zeta_{2}) + y^4 (-8 + 6 \zeta_{2}) + y^3 (-10 + 9 \zeta_{2})) \nonumber \\ &
    - 2 x^2 (-36 + 2 y^2 (-12 + \zeta_{2}) + 8 \zeta_{2} + y^4 (-8 + 6 \zeta_{2}) + 4 y (-13 + 7 \zeta_{2}) + y^3 (-10  \nonumber \\ &
    + 9 \zeta_{2})) + x^5 (90 + y^3 (8 - 28 \zeta_{2}) - 32 y (-2 + \zeta_{2}) - 20 \zeta_{2} - 5 y^4 (-4 + 3 \zeta_{2}) + 5 y^2 \nonumber \\ &
    (2 + 9 \zeta_{2})) + x^3 (-90 + 32 y (-2 + \zeta_{2}) + 20 \zeta_{2} - 4 y^3 (2 + \zeta_{2}) + 5 y^4 (-4 + 3 \zeta_{2}) +  \nonumber \\ & 
    y^2 (-10 + 11 \zeta_{2})) + x (-18 + 4 \zeta_{2} + 4 y^3 (-4 + 3 \zeta_{2}) + y^4 (-4 + 3 \zeta_{2}) + 8 y (-9 + 4 \zeta_{2})  \nonumber \\ &
    + y^2 (-50 + 39 \zeta_{2}))) + i \big( 2 \pi (y (24 + 20 y + 15 y^2 + 3 y^3) + x^2 y (24 + 20 y + 15 y^2 +  \nonumber \\ &
    3 y^3) +  x (-24 + 5 y^2 + 13 y^3 - 3 y^4 - 3 y^5))  \big) \big] + \big[\log(x) - 2 \log(1+x)\big] \big\{ \frac{1}{(-1 + x)^4 x y^2} \nonumber \\ &
    8 (3 y +   3 x^6 y  - x (3 + 6 y + 7 y^2) - x^5 (3 + 6 y + 7 y^2) + x^2 (12 + 19 y + 2 y^3) + x^4 (12 \nonumber \\ &
    + 19 y + 2 y^3) + 2 x^3 (-9 - 10 y - 5 y^2 + 4 y^3)) + i \big( \frac{1}{x y^2} 16 \pi (y + x^2 y - x (2 + y^2)) \big) \big\}  + \nonumber \\ &
    \log(y) \big\{ -\frac{1}{x y^2 (1 + y)} 8 (y (3 - y + 2 i \pi (1 + y)) + x^2 y (3 - y + 2 i \pi (1 + y)) - x (3  \nonumber \\ &
    - 3 y + y^2 - y^3 + 2 i \pi (2 + 2 y + y^2 + y^3)))  \big\} + \Big[ \log(1+y) - \frac{1}{4 i \pi} \log(y)^2 + \frac{1}{2 i \pi} \times \nonumber \\ &
    \log(1 +y)^2 + \frac{1}{i \pi}Li_{2}(\frac{1}{1+y}) +  \frac{1}{2 i \pi}\log(x)\log(y) - \frac{1}{i \pi} \log(1+x)\log(y) \Big] \Big( \frac{32 i \pi}{x y^2} \times \nonumber \\ &
    (y + x^2 y - x (2 + y^2)) \Big) +  \big[ \log(x)^2  + 4 Li_{2}(-x) \big] \big\{\frac{-1}{(-1 + x)^5 x (1 + x) y} (32 (1 + \nonumber \\ &
    x^8 + x^2 (2 - 4 y) + x^6 (2 - 4 y) - x y - x^7 y + x^3 y (-7 + 4 y) + x^5 y (-7 + 4 y) + \nonumber \\ &
    x^4 (6 + 4 y^2))) \big\}
\end{align}
\begin{align}
    \mathcal{J}_{1}^{fin} = & \frac{-1}{3 (-1 + x)^3 x^2 (1 + x) y (1 + y)^2} \Big[ m_{w}^2 (8 \pi^2 x^4 y (1 + y)^2 (4 + 3 y) + 
    x^8 y (6 i \pi (20 + 31 y \nonumber \\ &  
    + 18 y^2 + 3 y^3) - (1 + y) (5 \pi^2 (4 + 5 y + y^2) + 6 (4 (-9 + \zeta_{2}) + 5 y (-8 + \zeta_{2}) + y^2 ( \nonumber \\ & 
    -8 + \zeta_{2})))) + y (-6 i \pi (20 + 31 y + 18 y^2 + 3 y^3) + (1 + y) (5 \pi^2 (4 + 5 y + y^2) + 6 \times \nonumber \\ &
    (4 (-9 + \zeta_{2}) + 5 y (-8 + \zeta_{2}) + y^2 (-8 + \zeta_{2})))) + x^3 (6 i \pi (-144 - 227 y - 156 y^2 -  \nonumber \\ & 60 y^3 - 8 y^4 + 3 y^5) - (1 + y) (\pi^2 (-96 - 79 y - 29 y^2 - 41 y^3 + 5 y^4) + 6 (272 + y ( \nonumber \\ & 286 - 35 \zeta_{2}) + y^3 (44 - 5 \zeta_{2}) - 9 y^2 (-10 + \zeta_{2}) + y^4 (-8 + \zeta_{2}) - 32 \zeta_{2}))) + x^5 (6 i \pi \times \nonumber \\ &
    (144 + 227 y + 156 y^2 + 60 y^3 + 8 y^4 - 3 y^5) + (1 + y) (\pi^2 (-96 - 87 y + 11 y^2 + 7 y^3 \nonumber \\ &
    + 5 y^4) + 6 (272 + y ( 286 - 35 \zeta_{2}) + y^3 (44 - 5 \zeta_{2}) - 9 y^2 (-10 + \zeta_{2}) + y^4 (-8 + \zeta_{2}) \nonumber \\ & 
    - 32 \zeta_{2}))) - 2 x^2 (6 i \pi (-9 - 49 y - 61 y^2 - 24 y^3 + 4 y^4 + 3 y^5) - (1 + y) (\pi^2 (-6 - \nonumber \\ &
    57 y - 56 y^2 + 5 y^4) + 6 (17 + y^3 + y (79 - 9 \zeta_{2}) + y^2 (67 - 8 \zeta_{2}) + y^4 (-8 + \zeta_{2}) - \nonumber \\ &
    2 \zeta_{2}))) + 2 x^6 (6 i \pi (-9 - 49 y - 61 y^2 - 24 y^3 + 4 y^4 + 3 y^5) - (1 + y) (\pi^2 (-6 - 17 y \nonumber \\ &
    - 12 y^2 + 4 y^3 + 5 y^4) + 6 (17 + y^3 + y (79 - 9 \zeta_{2}) + y^2 (67 - 8 \zeta_{2}) + y^4 (-8 + \zeta_{2})  \nonumber \\ & 
    - 2 \zeta_{2}))) + x^7 (-6 i \pi (36 + 37 y + 12 y^2 + 12 y^3 + 12 y^4 + 3 y^5) + (1 + y) (\pi^2 (24 + y \nonumber \\ & 
    - 13 y^2 + 15 y^3 + 5 y^4) + 6 (-68 + 3 y^3 (-8 + \zeta_{2}) + y^4 (-8 + \zeta_{2}) - y^2 (-6 + \zeta_{2}) + \nonumber \\ &
    8 \zeta_{2} + y (-38 + 5 \zeta_{2})))) + x (6 i \pi (36 + 37 y + 12 y^2 + 12 y^3 + 12 y^4 + 3 y^5) - (1 + y)  \nonumber \\ & 
    (\pi^2 (24 + 9 y - 5 y^2 + 15 y^3 + 5 y^4) + 6 (-68 + 3 y^3 (-8 + \zeta_{2}) + y^4 (-8 + \zeta_{2}) - y^2 ( \nonumber \\  &
    -6 + \zeta_{2}) + 8 \zeta_{2} + y (-38 + 5 \zeta_{2}))))) \Big] + \Big[ \log(x) - 2 \log(1 +x)\Big] \Big\{ \frac{-1}{(-1 + x)^2 x y)} \nonumber \\ &
    4 m_{w}^2 (3 (2 + y) + 3 x^4 (2 + y) + x (3 + 10 y - y^2) + x^3 (3 + 10 y - y^2) - 2 x^2 (9 + 5 y \nonumber\\ &
    + 7 y^2) + 2 i \pi (-1 + x)^2 (4 + y + x^2 (4 + y) + x (10 + 4 y + y^2))) \Big\} +   \log(y) \Big\{  \nonumber \\ & 
    \frac{1}{(x^2 y (1 + y)^2)} 4 m_{w}^2 (-2 y (2 + y) - 2 x^4 y (2 + y) + x (6 - 9 y - 4 y^2 + 3 y^3 + 2 i \pi (1  \nonumber \\ &
   +  y)^2 (4 + y)) + x^3 (6 - 9 y - 4 y^2 + 3 y^3 + 2 i \pi (1 + y)^2 (4 + y)) + x^2 (15 - 2 y  \nonumber \\ &
   + 2 y^2 + 6 y^3 - y^4 + 2 i \pi (1 + y)^2 (10 + 4 y + y^2))) \Big\} + \Big[ \log(1+y) - \frac{1}{4 i \pi} \log(y)^2 \nonumber \\ &
   + \frac{1}{2 i \pi} \log(1 + y)^2  + \frac{1}{i \pi} Li_{2}\Big(\frac{1}{1+y}\Big) + \frac{1}{2 i \pi}\log(x)\log(y) - \frac{1}{i \pi} \log(1+x)\log(y)  \Big] \nonumber \\ &
   \Big\{ \frac{-16 i m_{w}^2 \pi}{x y}(4 + y + x^2 (4 + y) + x (10 + 4 y + y^2)) \Big\} + \Big[ \log(x)^2 + 4 Li_{2}(-x) \Big] \Big\{ \nonumber \\ & \frac{4 m_{w}^2}{(-1 + x)^3 x (1 + x)} (1 + x^6 + x^2 (1 - 6 y) + x^4 (1 - 6 y) + x (10 + y) + x^5 (10 + y) \nonumber \\ & - 2 x^3 (4 + 3 y)) \Big\}
\end{align}

\begin{align}
    \mathcal{K}_{1}^{fin}  = & \frac{1}{x^3} 2 m_{w}^4 \Big(4 + y + x^6 (4 + y) + x^2 (-4 + 11 y) + 
   x^4 (-4 + 11 y) - x (-7 + y^2) - \nonumber \\ & 
   x^5 (-7 + y^2) - 2 x^3 (13 + 5 y^2)) (-4 + 3 \zeta_{2} \Big)
\end{align}

Similar to Eq. \ref{eq:dquarkBORN} we can write the amplitudes for the $d$-type quark process.
\begin{align}
   \mathcal{F}^{d}_{1}(x,z) =  \mathcal{F}^{u}_{1}(x,y) , \quad
   \mathcal{J}^{d}_{1}(x,z) = -\mathcal{J}^{u}_{1}(x,y) , \quad
   \mathcal{K}^{d}_{1}(x,z) =  \mathcal{K}^{u}_{1}(x,y) .
\end{align}

The expressions for both $\mathcal{M}_{11}$ and $\mathcal{M}_{02}$ are too large in size to be given. However, the same can be provided upon request.
The process-independent universal resum exponent defined in \eq{eq:gn} which 
are used for DY-type processes can be found in the appendix of Ref.~\cite{Banerjee:2024xdh}.
\bibliographystyle{JHEP}
 \bibliography{qqbWW}
\end{document}